\documentclass[aps,amssymb,showpacs,floatfix,twocolumn,prx,reprint,longbibliography]{revtex4-2}
\usepackage{amssymb}
\usepackage{graphicx}
\usepackage{epsf}
\usepackage{amsmath}
\usepackage{xcolor}
\usepackage{romannum}
\usepackage{tikz}
\usepackage[colorlinks=true,citecolor=blue,linkcolor=blue,hypertexnames=false]{hyperref}
\usepackage{array}
\usepackage[normalem]{ulem}

\newcommand{\bOmega}{{\bf \Omega}}
\newcommand{\bk}{{\bf k}}

\newcommand{\bj}{{\bf j}}

\newcommand{\bE}{{\bf E}}
\newcommand{\bA}{{\bf A}}

\newcommand{\bB}{{\bf B}}
\newcommand{\br}{{\bf r}}

\newcommand{\ve}{{\varepsilon}}
\def \be{\begin{equation}}
\def \ee{\end{equation}}

\begin{document}
\pagenumbering{arabic}

\title{Magnetic field induces giant nonlinear optical response in Weyl semimetals}

\author{Grigory Bednik}
 \thanks{gbednik@unomaha.edu}
\affiliation{Department of Physics, University of Nebraska-Omaha, Omaha, Nebraska 68182, USA}

\author{Vladyslav Kozii}
\thanks{vkozii@andrew.cmu.edu}
\affiliation{Department of Physics, Carnegie Mellon University, Pittsburgh, Pennsylvania 15213, USA}

\begin{center}
\begin{abstract}
\newcounter{TypeOne}
\newcounter{TypeTwo}
\newcounter{TypeThree}

\setcounter{TypeOne}{1}
\setcounter{TypeTwo}{2}
\setcounter{TypeThree}{3}
We study the second-order optical response of Weyl semimetals in the presence of a magnetic field. We consider an idealized model of a perfectly linear Weyl node and use the Kubo formula at zero temperature to calculate the intrinsic contribution to photocurrent and second harmonic generation conductivity components. We obtain exact analytical expressions applicable at arbitrary values of frequency, chemical potential, and magnetic field. Our results show that finite magnetic field significantly enhances the nonlinear optical response in semimetals, while magnetic resonances lead to divergences in nonlinear conductivity. In realistic systems, these singularities are regularized by a finite scattering rate, but result in pronounced peaks which can be detected experimentally, provided the system is clean and interactions are weak. We also perform a semiclassical calculation that complements and confirms our microscopic results at small magnetic fields and frequencies.
\end{abstract}
\end{center}

\maketitle

\section{Introduction}

Recently, topological semimetals have been proposed as a promising platform for applications in solar cells and infrared photodetectors~\cite{NagaosaSolarCells2017,photodetection2020}. In contrast to traditional semiconductors, semimetals possess a gapless spectrum, allowing them to detect radiation in the terahertz range. Moreover, the geometry and topology of their band structure, stemming from the nonzero Berry curvature and orbital magnetic moments of the electrons, greatly increase their nonlinear optical response. These unique optoelectronic properties make semimetals strong candidates to outperform existing conventional solar cells and photodetectors.

The mechanism for generating large photocurrents in semimetals is the bulk photogalvanic effect (PGE). This term stands for the generation of rectified second-order  current under uniform irradiation of light in materials with broken inversion symmetry.  From the fundamental perspective, photocurrent and corresponding nonlinear conductivity are intimately related to the topology and geometry of the band structure~\cite{Morimotoe1501524,AhnNagaosa2020,Orensteinreview2021}. For example, it has been demonstrated that circular photogalvanic effect (CPGE), i.e., PGE with the circularly polarized light, is nearly quantized in Weyl and multi-Weyl materials, and is proportional to the monopole strength of a Weyl node~\cite{deJuan2017,Grushin2018,Reeseaba0509}. This connection makes nonlinear optics a powerful tool to probe band structure topology, motivating  intensive theoretical and experimental study in recent years~\cite{Rostamis2018,Ji2019,Ma2017,Osterhoudt2019,Wu2017,Haibin2018,Zyuzin2017,Golub2017}.

Depending on the microscopic origin of  photocurrent, one can distinguish intrinsic and extrinsic mechanisms. The former contribution is associated with photogeneration of electron-hole pairs, and can be calculated directly from the Bloch wave functions of a crystal. The latter mechanism, on the contrary, originates from the scattering processes or recombination of the electron-hole pairs. It depends on the details of the electron-electron and electron-phonon interactions and the nature of disorder.

Inspired by the potential applications of topological Weyl semimetals, we explore how a finite magnetic field may significantly enhance their nonlinear optical response~\cite{GolubIvchenko2018,Yao_2013,Belyanin2012,Ornigotti_2023}. To that end, we analyze a simple model of linear, nondegenerate band crossing point in three spatial dimensions, which is valid when the chemical potential and typical magnetic energy lie near the Weyl point. We calculate the intrinsic contribution to  nonlinear optical conductivity, which is expected to be dominant  in clean samples where interactions are weak. We additionally assume the breaking of inversion and all mirror symmetries, since the contributions of topological nodes with opposite chiralities would otherwise cancel each other out~\cite{deJuan2017}.

We apply the Kubo formula to calculate zero-temperature second-order conductivity tensors for both photocurrent and second harmonic generation (SHG), which is the response of the electric current at a double frequency of the incident electromagnetic wave. As such, we derive the most generic analytic expressions applicable for any frequency, chemical potential, and magnetic field, provided all the corresponding energy scales are smaller than the characteristic cutoff beyond which the spectrum cannot be considered as linear any longer. In the absence of any magnetic field, the second-order response of an isolated Weyl node is limited to a nonzero CPGE only, due to high emergent symmetry of a system at low energies~\cite{Avdoshkin2020}. Finite magnetic field reduces this symmetry, leading to a nonzero linear photogalvanic effect (LPGE) and SHG.

We find that even relatively weak fields may significantly increase nonlinear optical response. We analyze various limiting cases, particularly examining divergences in SHG originating from magnetic resonances. In a realistic experimental setup, these singularities would be smoothed out by a finite scattering rate; however, they could still substantially enhance nonlinear conductivity if interactions and disorder are weak. To corroborate our findings, we also perform a semiclassical calculation which agrees with the Kubo formula result in the limit of low fields and frequencies.

The remainder of the paper is organized as follows. In Section~\ref{Sec:Model}, we introduce the model and outline main steps of our calculation. In Sections~\ref{Sec:ShiftCurrent} and~\ref{Sec:SHG}, we derive the general expressions for the photocurrent and SHG conductivity components, respectively, and discuss  various limiting cases along with the divergences. The semiclassical calculation is presented in Section~\ref{Sec:semiclassics}. We wrap up with the discussion in Section~\ref{Sec:Discussion}. Technical details of our calculations are delegated to numerous Appendixes.

\begin{figure}
\centering
\includegraphics[width=.9\columnwidth]{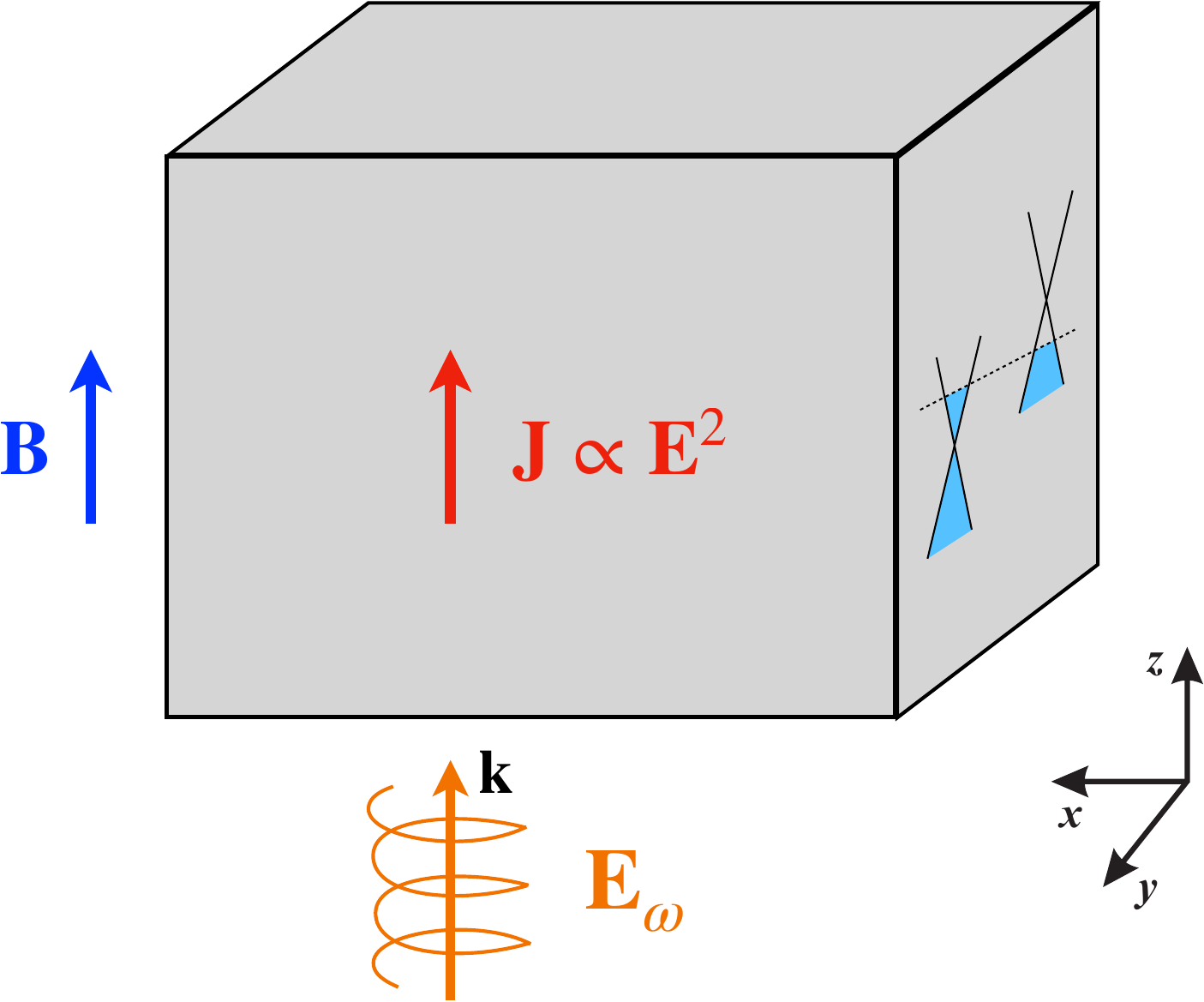}
\caption{Schematic picture of a setup considered in this paper. Weyl semimetal with nodes at different energies is placed under external static magnetic field $\bB$ along the $z-$direction. As an example, PGE experiment is shown: An electromagnetic wave with frequency $\omega$ propagates toward the sample along the magnetic field, and its electric field components lie in the $x-y$ plane. As a result, the photocurrent $\bf J$ flows in the $z-$direction.
}
\label{DevicePicture}
\end{figure}

\section{Model and Kubo formula}
\label{Sec:Model}
\begin{figure}
\centering
\includegraphics[width=1.\columnwidth]{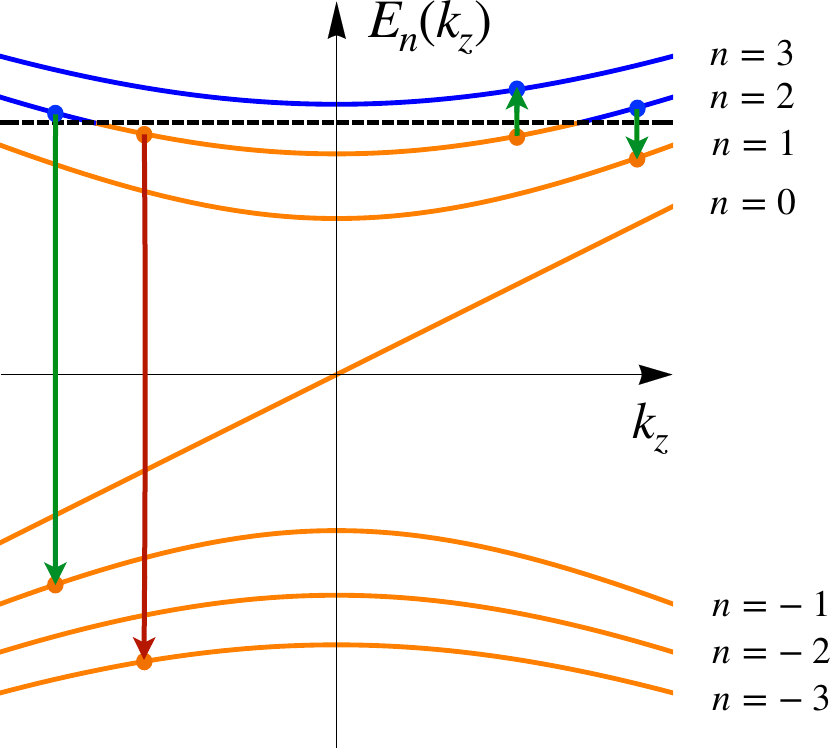}
\caption{A schematic picture of Landau levels around a single Weyl node, Eq.~\eqref{Eq:LL}. Black dashed line indicates chemical potential $\mu$, separating filled states (orange) from empty states (blue). Nonzero $\beta^{xyz}$ requires that the condition in Eq.~\eqref{Eq:region1} is satisfied, choosing the frequency range where the transitions between level $n$ crossing chemical potential and levels $n \pm 1,$ $-n+1$ (with $n=2$ in this figure) are allowed, as indicated by the green arrows. Nonzero $\beta^{xxz}$ additionally restricts frequencies according to Eq.~\eqref{Eq:region2}, which forbids the transition between levels $n$ and $-n-1$, as shown by the red arrow. See discussion after Eq.~\eqref{Eq:betaxxzsmallB} for more details. 
}
\label{LandauLevelsPicture}
\end{figure}

We perform our analysis for the simplest  model consisting of a single  Weyl node with linear dispersion. To obtain a physical result, we must sum up the contributions from all the nodes located close to the Fermi energy. Assuming magnetic field $B$ is in the $z-$direction (see Fig.~\ref{DevicePicture}) and using the Landau gauge $\bA = (- y B,0,0)$, we write the Hamiltonian as
\be  
 \hat H = \eta v_F \hbar \left[ \left( k_x - \frac{eB}{\hbar} y \right) \sigma_x -i\frac{\partial}{\partial y} \sigma_y + k_z \sigma_z \right]. \label{InitHam}
\ee 
Here $v_F$ is the Fermi velocity near the Weyl point,  $\eta = \pm 1$ is its chirality, $\sigma_{x,y,z}$ are the Pauli matrices, and $-e$ is the electron charge (we use the convention $e>0$, $B>0$). In this gauge, crystal momentum components $\hbar k_{x,z}$ remain good quantum numbers, and the energy levels are given by 
\be  
E_n(k_z) = \left\{ \begin{array}{lcr} \text{sgn}(n) \eta \hbar \sqrt{ v_F^2 k_z^2 + \omega_B^2 |n|},   & &n \ne 0, \\ -\eta \hbar v_F k_z, & &n=0, \end{array}  \right. \label{Eq:LL}
\ee 
where we introduced characteristic magnetic frequency $\omega_B^2 = 2 e B  v_F^2/\hbar$,  and $E_0(k_z)$ is a single chiral band with linear dispersion (see Fig.~\ref{LandauLevelsPicture}). The eigenfunctions can conventionally be expressed in terms of the Hermite polynomials; we present them along with the relevant matrix elements explicitly in Appendix~\ref{App:LL}.

Such an idealized description is valid as long as all the relevant energy scales, such as Fermi energy, magnetic energy, and frequency of external light, remain much smaller than the ultraviolet cutoff (of the order of the bandwidth) within which the dispersion may be approximated as linear. Furthermore, in Weyl semimetals, the nodes always come in pairs of opposite chirality. As we demonstrate below, the second-order response for a single node is proportional to its chirality; hence, the resulting current may only be nonzero if the nodes of opposite chirality are located at different energies. To achieve this, inversion and all mirror symmetries of the crystal must be broken, since these spatial symmetries relate the nodes with the opposite chiralities~\cite{deJuan2017}. Interestingly, to obtain a nonzero result, time-reversal symmetry may be preserved. While it guarantees the existence of another Weyl node at the same energy located symmetrically in the Brillouin zone, this node would have the same chirality and, hence, does not cancel out the total second-order current.

The most generic form of the second-order current is given in terms of the nonlinear conductivity tensor $\sigma^{\alpha \beta \gamma}(\omega_1,\omega_2)$ and reads as
\be  \label{Eq:2ndOhm}
j^{\gamma}(\Omega) = \sigma^{\alpha \beta \gamma}(\omega_1,\omega_2)   E^\alpha(\omega_1) E^\beta(\omega_2),
\ee
where we defined $\Omega \equiv \omega_1 + \omega_2$, $E^{\alpha}(\omega_i)$ are the components of the electric field at frequency $\omega_i$,  and the summation over the repeated indices $\alpha$, $\beta$ is implied. To calculate $\sigma^{\alpha \beta \gamma}(\omega_1,\omega_2)$, we use the generalization of the Kubo formula for the higher-order response functions. The diagrammatic derivation for the most general system within the Keldysh formalism can be found, e.g., in Ref.~\cite{JoaoLopes2018}. In this work, however, we find it more convenient to use Matsubara formalism and follow the steps of Ref.~\cite{Avdoshkin2020}. In particular, we express the second-order conductivity through the three-point correlation function $\chi^{\alpha \beta \gamma}$:
\be
\sigma^{\alpha \beta \gamma}(i\omega_1,i\omega_2)=\frac{\chi^{\alpha \beta \gamma}(i\omega_1, i\omega_2) +  \chi^{\beta \alpha  \gamma}(i\omega_2, i\omega_1) }{\omega_1 \omega_2}, \label{Eq:sigmachi}
\ee 
where in the zero-temperature limit
\begin{align}  
&\chi^{\alpha \beta \gamma}(i\omega_1,i\omega_2) = \frac1V \int\frac{d\ve}{2\pi} \text{Tr}\left[ \hat j^\alpha G(i\ve - i\omega_1)    \times \right. \nonumber \\ &\left. \times   \hat j^\beta G(i\ve - i\Omega) \hat j^\gamma G(i\ve) \right], \label{Eq:chi}
\end{align}
and $V$ is the system's volume. In this expression, the trace also implies integration over the intermediate coordinates, and we introduced the exact Green's function in the external uniform magnetic field:
\be  
G(i\ve, \br_1, \br_2) = \sum_{n, k_x,k_z}\frac{|\Psi_{n,k_x,k_z}(\br_1)\rangle \langle \Psi_{n,k_x,k_z}(\br_2)|}{i\ve - E_n(k_z) + \mu}.
\ee 
Here, $|\Psi_{n,k_x,k_z}(\br)\rangle$ are the eigenstates of the Hamiltonian in Eq.~\eqref{InitHam} with quantum numbers $n$, $k_x$, and $k_z$, presented in Appendix~\ref{App:LL}, and $\mu$ is the chemical potential calculated with respect to the position of the node. The current operator is given by 
\be
\hat j^{\alpha} = - \frac{e}{\hbar} \frac{\delta H_{\bk}}{\delta k^\alpha} = - \eta e v_F \sigma^{\alpha},
\ee
where $H_{\bk}$ is the band Hamiltonian. We stress that all the higher-order derivatives vanish for a Weyl node, $\delta^2 H_{\bk}/\delta k^\alpha \delta k^\beta = 0$, consequently, Eq.~\eqref{Eq:chi} is the only contribution to the second-order conductivity tensor. Finally, the factor $1/\omega_1 \omega_2$ in Eq.~\eqref{Eq:sigmachi} originates from the relation between the electric field and homogeneous but time/frequency-dependent part of the vector potential, $\bE(\omega) = i\omega \bA(\omega)$.

When rewritten in momentum space, the trace in Eq.~\eqref{Eq:chi} implies summation over quasimomenta $k_x$ and $k_z$ in addition to the matrix trace in pseudospin ($\boldsymbol \sigma$) space. The summation over $k_x$ merely accounts for the degeneracy of each Landau level (at any given $k_z$) and results in the factor $\sum_{k_x} \to eBS/2\pi\hbar$, where $S$ is the area of the system in the $x-y$ plane. Integration over $\ve$ can be performed explicitly resulting in
\begin{align}
&\chi^{\alpha \beta \gamma}(i\omega_1, i\omega_2) = (e v_F)^3 \frac{\eta eB}{2 \pi \hbar} \int_{-\infty}^{\infty} \frac{d k_z}{2\pi} \sum_{n_1, n_2, n_3} Z_{n_1 n_2 n_3}^{\alpha \beta \gamma} \times \nonumber \\ &\times \frac1{i \hbar \omega_1 + \ve_1 - \ve_3} \left[ \frac{\Theta(\ve_2) - \Theta(\ve_1)}{i \hbar \omega_2 + \ve_2 - \ve_1}     -    \frac{\Theta(\ve_2) - \Theta(\ve_3)}{i \hbar\Omega + \ve_2 - \ve_3}   \right], \label{2ndOrderGeneric}
\end{align}
where $\Theta(\ve)$ is the Heaviside step function, and we defined  $\ve_i = E_{n_i}(k_z) - \mu$ and 
\be 
Z_{n_1 n_2 n_3}^{\alpha \beta \gamma} = \langle \Psi_{n_3} | \sigma^\alpha|\Psi_{n_1}  \rangle \langle \Psi_{n_1} | \sigma^\beta|\Psi_{n_2}  \rangle \langle \Psi_{n_2} | \sigma^\gamma|\Psi_{n_3}  \rangle. \label{Eq:Zs}
\ee 
All the matrix elements entering Eq.~\eqref{Eq:Zs} are calculated explicitly in Appendix~\ref{App:LL}. In order to obtain physical conductivity, we will perform analytic continuation $i\omega_{1,2} \to \omega_{1,2} + i0$ from the upper complex half-plane to the real frequencies axis. 

Equation (\ref{2ndOrderGeneric}) is the starting point for calculating photocurrent and second harmonic generation, which we discuss in detail in the following sections.  As a check, we reproduce the exact answer for $\chi^{\alpha \beta \gamma}(i\omega_1, i\omega_2)$ in the limit $B\to 0$, obtained in Ref.~\cite{Avdoshkin2020}, in Appendix~\ref{App:B=0}. We note that, in this limit, only CPGE (with $\omega_2 = -\omega_1$) remains nonzero for an ideal Weyl node considered in this paper, while LPGE and SHG (with $\omega_1 = \omega_2$) vanish. As we demonstrate below, the presence of a finite magnetic field, which reduces the symmetry of the system, leads to nonzero LPGE and SHG.

\section{Photocurrent}
\label{Sec:ShiftCurrent}

In this section, we consider the photogalvanic effect, i.e., generation of the second-order dc photocurrent under incident light. In particular, we calculate the injection current, which is the dc component that grows unrestricted with time in the clean noninteracting limit:
\be  
\frac{d j^{\gamma}}{d t} = \beta^{\alpha \beta \gamma}(\omega)   E^\alpha(\omega) E^\beta(-\omega).
\ee
In realistic systems, the injection current eventually saturates after scattering time $\tau$  to the value
$j^{\gamma} \approx \tau \beta^{\alpha \beta \gamma}(\omega)   E^\alpha(\omega) E^\beta(-\omega)$~\cite{deJuan2017,Levchenko2017}. However, if the system is weakly interacting and sufficiently clean, this contribution is the dominant one, making it the focus of this section. 

To extract injection current and calculate nonlinear conductivity $\beta^{\alpha \beta \gamma}$ from the general expressions in Eqs.~\eqref{Eq:sigmachi} and~\eqref{2ndOrderGeneric}, we perform analytic continuation $i\omega_1 \to \omega + \Omega + i0$, $i\omega_2 \to -\omega + i0$, where $\omega$ is the frequency of incident light and $\Omega \to 0$.  The photocurrent linearly growing in time is captured then by the   contribution to $\sigma^{\alpha \beta \gamma}(\omega+\Omega, -\omega)$ proportional to $1/\Omega$. Focusing on this contribution only as the leading one, we obtain after straightforward  calculation
\begin{widetext}
\be \label{Eq:injcurrent}
\sigma^{\alpha \beta \gamma}(\omega + \Omega, -\omega) = \frac{i(e v_F)^3 }{\Omega \omega^2} \cdot \frac{\eta eB}{\hbar^2}  \int_{-\infty}^{\infty} \frac{d k_z}{2\pi} \sum_{n_1, n_2} \left[ \Theta(\ve_1) - \Theta(\ve_2) \right] \left[\delta(\ve_2 - \ve_1 -\hbar\omega) Z_{n_1 n_2 n_2}^{\alpha \beta \gamma} + \delta(\ve_2 - \ve_1 +\hbar\omega) Z_{n_1 n_2 n_2}^{\beta \alpha \gamma} \right].
\ee
\end{widetext}
Tensor $\beta (\omega)$ can be found as 
\be \label{Eq:betagen}
\beta^{\alpha \beta \gamma}(\omega) = -i \Omega \sigma^{\alpha \beta \gamma}(\omega + \Omega, -\omega)
\ee
with $\Omega \to 0$.

The structure of the form-factors $Z^{\alpha \beta \gamma}$ and corresponding matrix elements from Eqs.~\eqref{Eq:Zs} and~\eqref{SMEq:matrixelements} dictate that the only nonzero components are those with $\gamma = z$, i.e., the direction of the injection current is along the magnetic field. Specifically, we find that the nonzero components for an ideal Weyl node are $\beta^{xyz} = -\beta^{yxz}$ and $\beta^{xxz} = \beta^{yyz}$. The  latter two components are nonzero only in the presence of a finite magnetic field.

This conclusion seems at first in contradiction with the exact answer for the $B=0$ case, derived in Appendix~\ref{App:B=0}, which is fully isotropic and obviously does not have any preferred direction. The seeming inconsistency originates from different orders of limits that do not commute. Indeed, to derive Eq.~\eqref{SMEq:chi_B=0} for $B=0$, we assumed that $\omega_B \propto \sqrt{B} \to 0$, while keeping $\omega_1$ and $\omega_2$ (hence, $\Omega$) most generic and finite. On the contrary, to derive Eq.~\eqref{Eq:injcurrent}, we performed analytic continuation and set $\Omega \to 0$ first, keeping magnetic field finite (though possibly small). Hence, the limits $\omega_B \to 0$ and $\Omega \to 0$ do not commute and lead to different answers, depending on the order. Physically, the order of limits is controlled by the dimensionless parameter $\omega_B \tau$. The regime $\omega_B \tau \to 0$ leads to Eq.~\eqref{SMEq:chi_B=0}, while the opposite limit, $\omega_B \tau \to \infty$, corresponds to Eq.~\eqref{Eq:injcurrent}, with $i/\Omega$ being replaced by $\tau$. We return to this issue when discussing semiclassical calculation in Section~\ref{Sec:semiclassics}.

The presence of the delta-function in Eq.~\eqref{Eq:injcurrent} allows us to perform the integration over momentum $k_z$ explicitly. After straightforward but lengthy calculation with the details presented in Appendix~\ref{App:injection}, we find the most general answer for the components of tensor $\beta(\omega)$: 
\begin{widetext}
\begin{align}
    &\beta^{xyz}(\omega) = -\beta^{yxz}(\omega) = - \frac{i \eta e^3}{4\pi \hbar^2} \frac{\omega_B^2 \text{sgn} (\omega)}{|\omega^4 - \omega_B^4|} \left[\Theta\left(\omega^2 + \omega_B^2 - 2 \left|\omega \frac{\mu}{\hbar}\right|\right)  + \text{sgn}\left(\omega^2 - \omega_B^2\right) \Theta\left(\left|\omega^2 - \omega_B^2\right| - 2\left|\omega \frac{\mu}{\hbar}\right| \right) \right]  \, \times  \nonumber\\ &\times \sum_{n=0}^{\infty} \text{Re}\sqrt{(\omega^2 - \omega_B^2)^2 -4\omega^2 \omega_B^2 n}, \label{Eq:betaxyz} \\     
        &\beta^{xxz}(\omega) = \beta^{yyz}(\omega) =  -\frac{\eta e^3}{4\pi \hbar^2}  \frac{\omega_B^2 \text{sgn} (\mu)}{|\omega^4 - \omega_B^4|} \left[\Theta\left(\omega^2 + \omega_B^2 - 2 \left|\omega \frac{\mu}{\hbar}\right|\right) - \Theta\left(|\omega^2 - \omega_B^2| - 2\left|\omega \frac{\mu}{\hbar}\right| \right)  \right]  \times \nonumber \\ &\times \sum_{n=0}^{\infty}\text{Re}\sqrt{(\omega^2 - \omega_B^2)^2 -4\omega^2 \omega_B^2 n}. \label{Eq:betaxxz}
\end{align}
\end{widetext}

The dependence of these components on frequency and magnetic field is illustrated in Figs.~\ref{Fig:injB} and~\ref{Fig:injw} (and Fig.~\ref{AppFig:densitypge} in Appendix~\ref{App:injection}). Two main effects of a finite magnetic field are nonzero components $\beta^{xxz} = \beta^{yyz}$ and a significant enhancement of the photocurrent in a certain parameter range as compared with the setup without field.  Our result agrees with the expression obtained by Golub and Ivchenko in Ref.~\cite{GolubIvchenko2018}. Below, we analyze different limiting cases in more detail.

\begin{figure}
  \centering
  \includegraphics[width=1.\columnwidth]{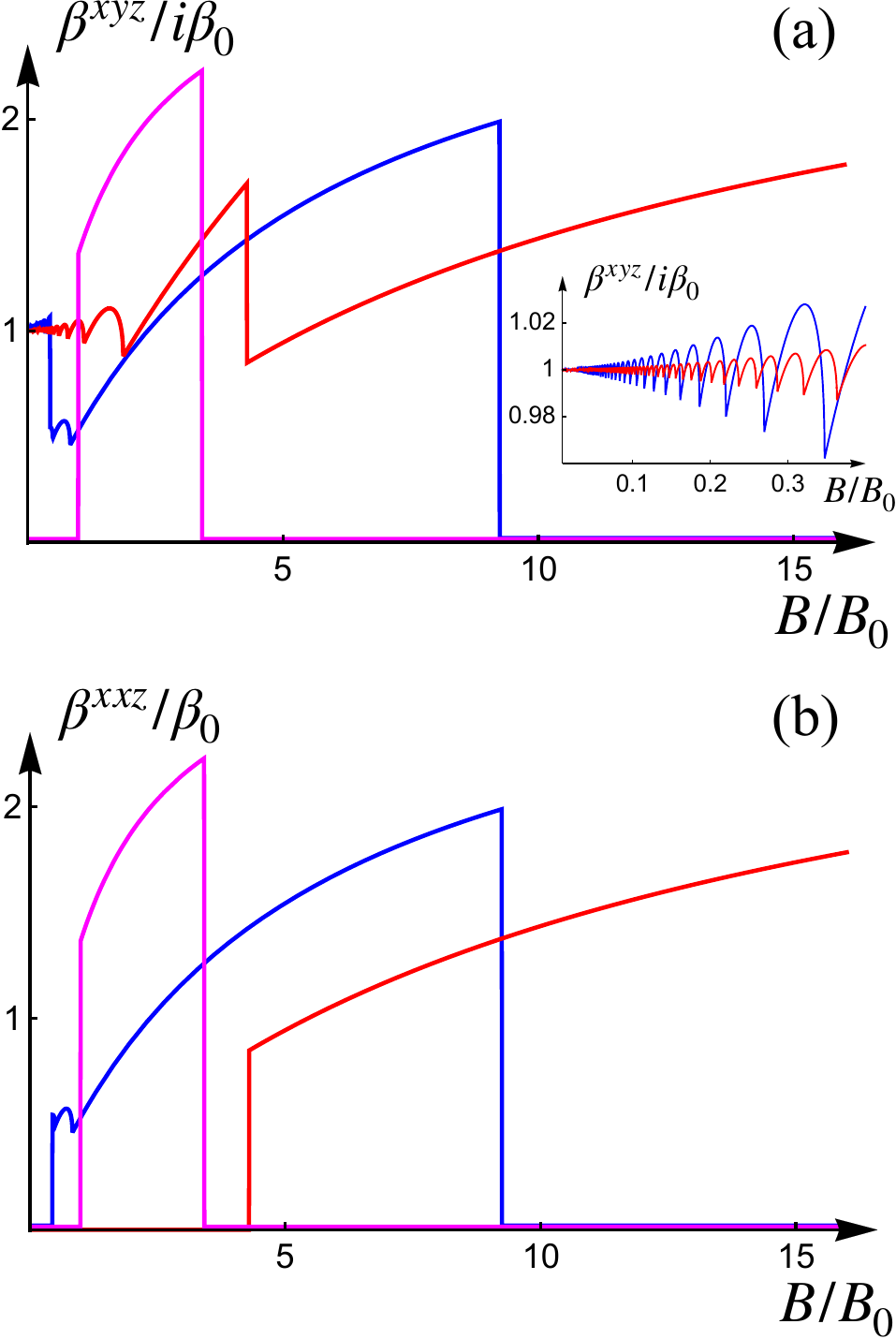}
    \caption{The components of the second-order dc conductivity $\beta^{xyz}$ (a)  and $\beta^{xxz}$ (b) as a function of magnetic field $B$, calculated from Eqs.~\eqref{Eq:betaxyz}--\eqref{Eq:betaxxz} at fixed chemical potential $\mu$. Different curves correspond to the ratio $\hbar \omega/\mu = 2.2$ (blue), 3.3 (red), and 1.1 (magenta). The component $\beta^{xyz}$ approaches $i \beta_0$ as $B \to 0$, provided $\hbar \omega > 2 \mu$. The inset in (a) shows quantum oscillations of $\beta^{xyz}$ at weak fields at $\hbar \omega > 2 \mu$. The regions where the different components are nonzero are given by the inequalities in Eqs.~\eqref{Eq:region1} and~\eqref{Eq:region2}.  We use the units $\beta_0 \equiv -\eta e^3/12 \pi \hbar^2$ and $B_0 \equiv \mu^2/2e\hbar v_F^2$.}
  \label{Fig:injB}
\end{figure}

\begin{figure}
  \centering
  \includegraphics[width=1.\columnwidth]{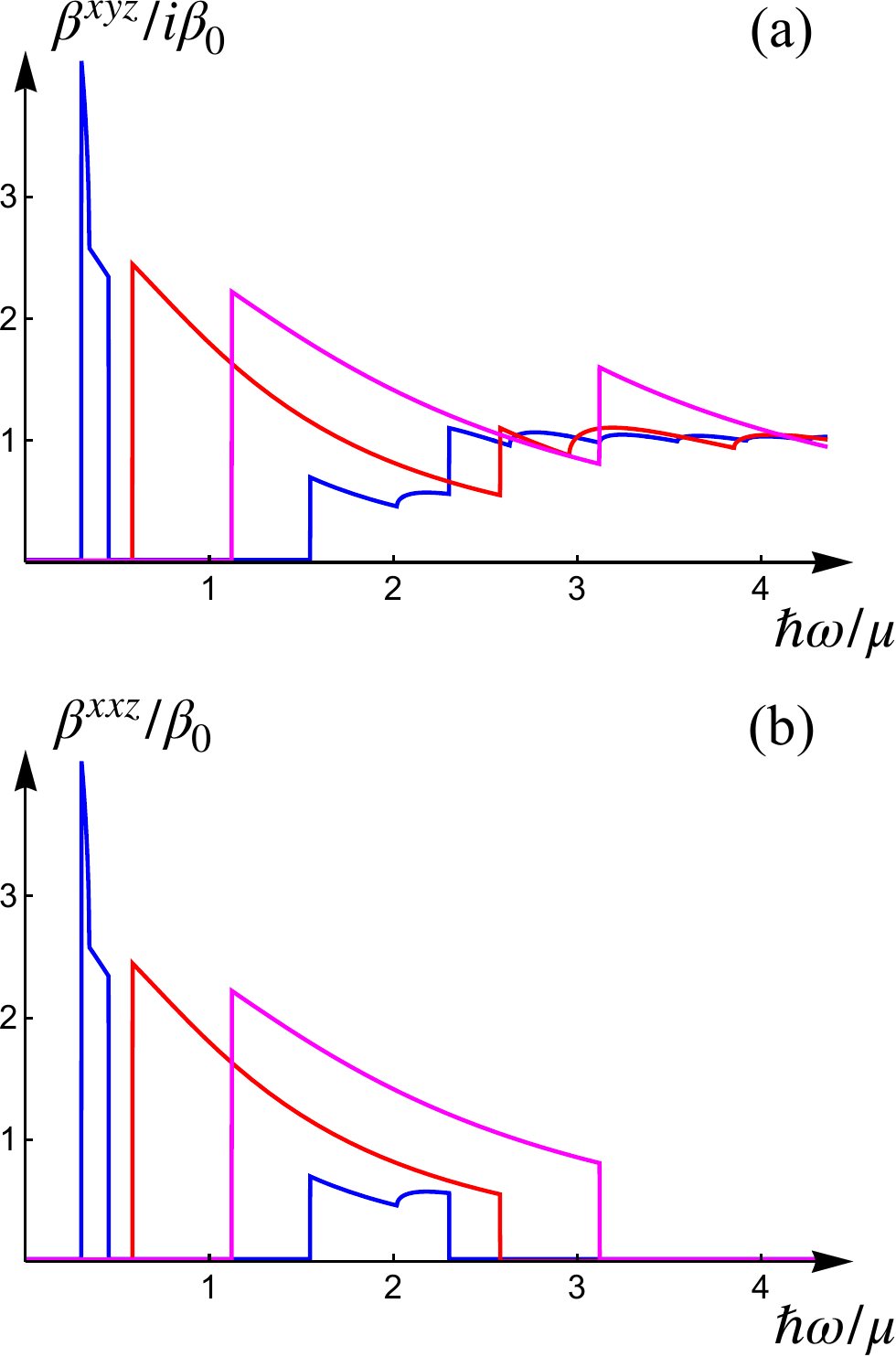}
    \caption{The components of the second-order dc conductivity $\beta^{xyz}$ (a)  and $\beta^{xxz}$ (b) as a function of the light frequency $\omega$, calculated from Eqs.~\eqref{Eq:betaxyz}--\eqref{Eq:betaxxz}. Different curves correspond to the fixed ratio $B/B_0 = \hbar^2 \omega_B^2/\mu^2  = 0.7$ (blue), 1.5 (red), and 3.5 (magenta). The blue peak at small frequencies is described by Eqs.~\eqref{Eq:dcpeakpos}--\eqref{Eq:dcpeakrange}. The blue ``bump'' around $\hbar \omega \approx 2\mu$ is described by Eq.~\eqref{Eq:betaxxzsmallB}. The component $\beta^{xyz}$ approaches $i \beta_0$ at high frequencies $\omega \gg \mu/\hbar$, $\omega_B$ in an oscillatory manner. The units $\beta_0$ and $B_0$ are the same as in Fig.~\ref{Fig:injB}. } 
  \label{Fig:injw}
\end{figure}

First, we consider component $\beta^{xyz}$ given by Eq.~\eqref{Eq:betaxyz}. This component describes circular photogalvanic effect (CPGE), i.e., the part of the photocurrent which changes sign upon switching light chirality. It can be shown from the structure of the step functions that this contribution is nonzero provided
\be  
|\hbar\omega_B^2 - 2\omega \mu| < \hbar\omega^2, \label{Eq:region1}
\ee
and we focus on the case $\mu, \omega>0$ for simplicity in this section henceforth.

Typical dependence on frequency and magnetic field (at fixed chemical potential) is shown in Figs.~\ref{Fig:injB}(a) and~\ref{Fig:injw}(a). In the limit of vanishing magnetic field $\omega_B \to 0$ and finite frequency, we reproduce the known quantized result~\cite{deJuan2017,Avdoshkin2020}:
\be  
\beta^{xyz}(\omega, \omega_B=0) = i \beta_0 \Theta(\hbar\omega - 2\mu), \quad  \beta_0 \equiv -\frac{\eta e^3}{12 \pi \hbar^2}.
\label{Eq:Quantized_result}
\ee 

We find that, upon finetuning, $\beta^{xyz}$ exhibits an unsaturated growth in the low-frequency low-field limit. Indeed, consider the limit
\be  
|\hbar\omega_B^2 - 2\omega \mu| < \hbar \omega^2 \ll \hbar\omega_B^2 \approx 2\omega \mu.  \label{Eq:dcpeakpos}
\ee  
At fixed magnetic field and chemical potential, this limit corresponds to the finetuned frequency $\omega \approx \hbar\omega_B^2 / 2\mu$ and implies the scale separation $\omega \ll \omega_B \ll 2\mu/\hbar$. We obtain then that, in this regime, CPGE grows as the frequency and magnetic field decrease simultaneously: 
\be  
\beta^{xyz}(\omega) \approx -i \frac{\eta e^3}{24 \pi \hbar^2} \frac{\omega_B^2}{\omega^2} \approx -i \frac{\eta e^3}{12 \pi \hbar^3}\frac{\mu}{\omega} = i \beta_0 \frac{\mu}{\hbar\omega}.
\ee 
We emphasize that such a giant photocurrent is possible due to  a finite though small magnetic field. This regime, however, is only realized in a very narrow window of frequencies: 
\be  
\left|\omega - \frac{\hbar\omega_B^2}{2\mu}\right| \lesssim \omega_B \left( \frac{\hbar\omega_B}{2\mu} \right)^3. \label{Eq:dcpeakrange}
\ee

Finally, we  comment on the field dependence shown in Fig.~\ref{Fig:injB}. It is calculated for the grand canonical ensemble, i.e., under the assumption that the chemical potential remains a constant. A more physically natural setup, however, implies a fixed number of particles, with the chemical potential being a function of the magnetic field, $\mu \to \mu(B)$, which must be determined self-consistently. We do not analyze this case in detail here, since it is very sensitive to a particular model. Indeed, as we discussed earlier, the Weyl nodes always come in pairs of different chiralities, and the nonzero second-order response requires these nodes to be located at different energies.  Node separation introduces at least one additional energy scale which makes reaching  any universal conclusions problematic. For any specific Weyl material with a particular location of the nodes and density, however, the dependence $\mu(B)$ can be easily evaluated numerically from the requirement of a constant particle number.

Next, we consider linear photogalvanic effect (LPGE) described by Eq.~\eqref{Eq:betaxxz}. Its dependence on frequency and field is shown in Figs.~\ref{Fig:injB}(b) and~\ref{Fig:injw}(b). Unlike CPGE, finite LPGE for an ideal Weyl node necessarily requires a finite magnetic field. Indeed, we see from Eq.~\eqref{Eq:betaxxz} that $\beta^{xxz}$ is nonzero only in the parameter range given by 
\be  
|\hbar\omega_B^2 - 2\omega \mu| < \hbar\omega^2 < \hbar\omega_B^2 + 2 \omega \mu. \label{Eq:region2}
\ee 
In this region, CPGE and LPGE components are related as 
\be  
\beta^{xxz}(\omega) = -i \beta^{xyz}(\omega). 
\ee 
In particular, LPGE demonstrates the same unsaturated growth at small frequencies and fields around $\omega \approx \hbar\omega_B^2/2\mu$. Additionally, in the limit of a vanishing field $\omega_B \to 0$, it is nonzero in the narrow frequency window of the width $\hbar\omega_B^2/\mu$ around $\omega \approx 2\mu/\hbar$ and equals 

\be  
\beta^{xxz}\left(\omega_B \ll \omega, \frac{\mu}{\hbar}\right) \approx \frac{\beta_0}2 \Theta \left(\frac{\omega_B^2}{\omega} - \left| \omega - 2\frac{\mu}{\hbar}\right|   \right). \label{Eq:betaxxzsmallB}
\ee

The conditions in Eqs.~\eqref{Eq:region1} and~\eqref{Eq:region2} have clear physical meaning. Indeed, consider any energy level $E_n(k_z)$ which crosses chemical potential $\mu > 0$ (we assume $n>0$ for simplicity). Equation $E_n(k_n) = \mu$ defines the effective ``Fermi momentum'' $k_n$ for the $n$-th Landau level. Consequently, the inequality in Eq.~\eqref{Eq:region1} can be rewritten as a combination of two requirements: 
\be 
1. \hbar\omega > E_{n+1}(k_n) - E_n(k_n) \, \Rightarrow \, \hbar \omega > \sqrt{\mu^2 + \hbar^2 \omega_B^2}- \mu \label{Ineq:1}
\ee
and 
\begin{align}
&2. \,\, \text{one of the two conditions}, \nonumber \\  &\hbar \omega < E_{n}(k_n) - E_{n-1}(k_n) \, \Rightarrow \, \hbar \omega < \mu - \sqrt{\mu^2 - \hbar^2 \omega_B^2} \nonumber
\end{align}
or 
\be 
\hbar \omega > E_{n}(k_n) - E_{-n+1}(k_n) \, \Rightarrow \, \hbar \omega > \mu + \sqrt{\mu^2 - \hbar^2 \omega_B^2},
\ee 
where we defined $E_{-n} \equiv - E_n$ for $n>0$. As is clear from Fig.~\ref{LandauLevelsPicture}, these requirements exactly imply that the frequency range is such that the direct transitions between the Landau levels $n \to n+1$ and $n \to n-1$ or $n \to -n+1$ are allowed by Pauli exclusion principle. In contrast, additional restriction imposed by the inequality in Eq.~\eqref{Eq:region2}, $\omega^2 < \omega_B^2 + 2 \omega \mu$, implies that 
\be
\hbar \omega < E_{n}(k_n) - E_{-n-1}(k_n) \, \Rightarrow \, \hbar \omega < \sqrt{\mu^2 + \hbar^2 \omega_B^2} + \mu. \label{Ineq:3}
\ee
This condition, in turn, picks the frequencies which forbid direct transitions $n \to -n-1$ due to the Pauli exclusion principle, see Fig.~\ref{LandauLevelsPicture}. As we discuss in the next Section, the frequencies at the boundaries of these regions define the resonances which lead to divergences in SHG.


\section{Second harmonic generation}
\label{Sec:SHG}

Another type of the second-order response generically present in noncentrosymmetric materials is second harmonic generation, a  frequency doubling of incident light through its interaction with media. To calculate the corresponding conductivity components, we start again with Eqs.~\eqref{Eq:sigmachi} and~\eqref{2ndOrderGeneric} and put $\omega_1 = \omega_2 = \omega$. This time, however, it is more convenient to do calculation  in Matsubara frequencies directly and perform the analytic continuation at the last step. The SHG components of conductivity are symmetric in the first two indices because the two frequencies coincide.  Furthermore, we note that, since the Kubo formula predicts the full response to a given field configuration, the left-hand side of Eq.~\eqref{Eq:sigmachi} should be understood as $\sigma^{\alpha \beta \gamma}(i\omega,i\omega) + \sigma^{\beta \alpha  \gamma}(i\omega,i\omega) = 2 \sigma^{\alpha \beta \gamma}(i\omega,i\omega)$.

Straightforward but lengthy calculation shows that the only nonzero SHG components for a Weyl node are  $\sigma^{xxz} = \sigma^{yyz}$, $\sigma^{zxx} = \sigma^{xzx} = \sigma^{zyy} = \sigma^{yzy}$, and $\sigma^{xzy} = \sigma^{zxy} = -\sigma^{yzx} = -\sigma^{zyx}$. Below, we consider all these components in detail separately.

\subsection{${\boldsymbol \sigma^{xxz}}$ component of SHG}

We start by considering the components $\sigma^{xxz} = \sigma^{yyz}$. The general expression in Matsubara frequencies is derived in Appendix~\ref{App:SHGxxz} and given by

\be  
\sigma^{xxz}(i\omega) \equiv \sigma^{xxz}(i\omega,i\omega) = 
\frac{\eta e^3\omega_B^2 \,\text{sgn} (\mu)}{32 \pi^2 \hbar^2 \omega^2} \sum_{n=0}^{N_{\max}}f_n(i\omega), \label{Eq:SHGgeneralxxz}
\ee
with 
\be  
f_n(i\omega) = \frac1{ia_n(i\omega)} \ln \frac{[ia_n(i\omega) + k_n][i a_n(i\omega) - k_{n+1}]}{[ia_n(i\omega) - k_n][i a_n(i\omega) + k_{n+1}]}. \label{Eq:fn}
\ee

We have defined the short-hand notations:
\begin{align}
N_{\max} &\equiv \left\lfloor\frac{\mu^2}{\hbar^2\omega_B^2}\right\rfloor, \qquad k_n \equiv \text{Re}\sqrt{\frac{\mu^2}{\hbar^2} - \omega_B^2 n},  \nonumber \\ a_n(i\omega) &\equiv \left[ \omega_B^2 n + \frac{(\omega_B^2 + \omega^2)^2}{4\omega^2} \right]^{1/2}, \label{Eq:Nmax}
\end{align}
such that $\hbar k_n/v_F$ is the momentum at which the chemical potential $\mu$ crosses the $n$-th Landau level and $\lfloor x \rfloor$ is the floor function, i.e., the greatest integer not exceeding $x$.

To extract physical response, we analytically continue Eq.~\eqref{Eq:SHGgeneralxxz} to real frequencies, $i\omega \to \omega + i 0$. The result as a function of frequency and magnetic field is shown in Fig.~\ref{Fig:2nd1} (and in Fig.~\ref{AppFig:dplot2wxxz} in Appendix~\ref{App:SHGxxz}). Unlike photocurrent, SHG components have both real and imaginary parts. To regularize the divergences that we discuss in more detail below, we keep small but finite imaginary part of frequency, which physically represents finite single-particle scattering rate. We emphasize again that we neglect the dependence of chemical potential $\mu$ on $B$ in Fig.~\ref{Fig:2nd1}(b), which corresponds to the great canonical ensemble and must be modified if the density of particles is fixed.

The analytic continuation in Eq.~\eqref{Eq:SHGgeneralxxz} can be performed explicitly; however, since the resulting expressions are rather cumbersome, we present them in Appendix~\ref{App:SHG}. Instead, here, we discuss the divergences and most interesting limiting cases.

We start with the case of sufficiently large magnetic field $\hbar \omega_B > |\mu|$. In this extreme quantum limit (EQL), only the $n=0$ term in Eq.~\eqref{Eq:SHGgeneralxxz} contributes, and we find 
\begin{align} \label{Eq:EQLxxz}
\sigma^{xxz}(\omega) &=\frac{\eta e^3 }{16 \pi^2 \hbar^2 }\frac{\omega_B^2}{\omega(\omega_B^2-\omega^2)} \left\{\ln \left| \frac{2\mu\omega - \hbar(\omega_B^2-\omega^2)}{2\mu\omega + \hbar(\omega_B^2-\omega^2)} \right| \right. \nonumber \\ &- \left. \pi i \, \text{sgn}(\mu) \Theta\left( 2|\omega\mu| - \hbar|\omega_B^2 - \omega^2|  \right)\right\}.
\end{align}
We see that the real part   diverges logarithmically at $\hbar\tilde \omega_1 = \pm (\sqrt{\hbar^2 \omega_B^2 + \mu^2}\pm \mu)$, while the imaginary part has power-law divergence at $\omega = \pm\omega_B$. The real part also exhibits divergence at $\omega = \pm\omega_B$ if small but finite scattering rate is taken into account [which is absent in Eq.~\eqref{Eq:EQLxxz}].

In the opposite limit of vanishing field $B\to 0$ ($\omega_B \to 0$), the summation in Eq.~\eqref{Eq:SHGgeneralxxz} can be approximated by an integral, leading to 
\begin{align}
\sigma^{xxz}(\omega) &\approx \frac{\eta e^3 \omega_B^2}{16 \pi^2 \hbar \omega^2} \left\{ \frac{4\mu}{4\mu^2 - \hbar^2\omega^2}    \right. \nonumber\\ &+ \left. \pi i \delta(\hbar|\omega| - 2|\mu|) \,\text{sgn}(\omega \mu) \right\}. \label{Eq:SHGsmallB}
\end{align}
This expression is valid provided $\hbar^2 \omega_B^2 \ll \hbar^2 \omega^2,$ $\mu^2$, $|\hbar^2 \omega^2 - 4\mu^2|$. We emphasize that its real part has more sophisticated behavior in the vicinity of $\hbar |\omega| \approx 2|\mu|$, while the width of the delta-function is proportional to the magnetic field. 

In the low-frequency limit, $\omega \to 0$, we may simply Taylor expand Eq.~\eqref{Eq:SHGgeneralxxz} and obtain
\be  \label{Eq:xxzloww}
\sigma^{xxz}(\omega) \approx - \frac{\eta e^3  \mu}{4 \pi^2 \hbar^3 \omega_B^2}. 
\ee 
We see that this component grows unlimited if additionally magnetic field approaches zero.

\begin{figure}
  \centering
  \includegraphics[width=1.\columnwidth]{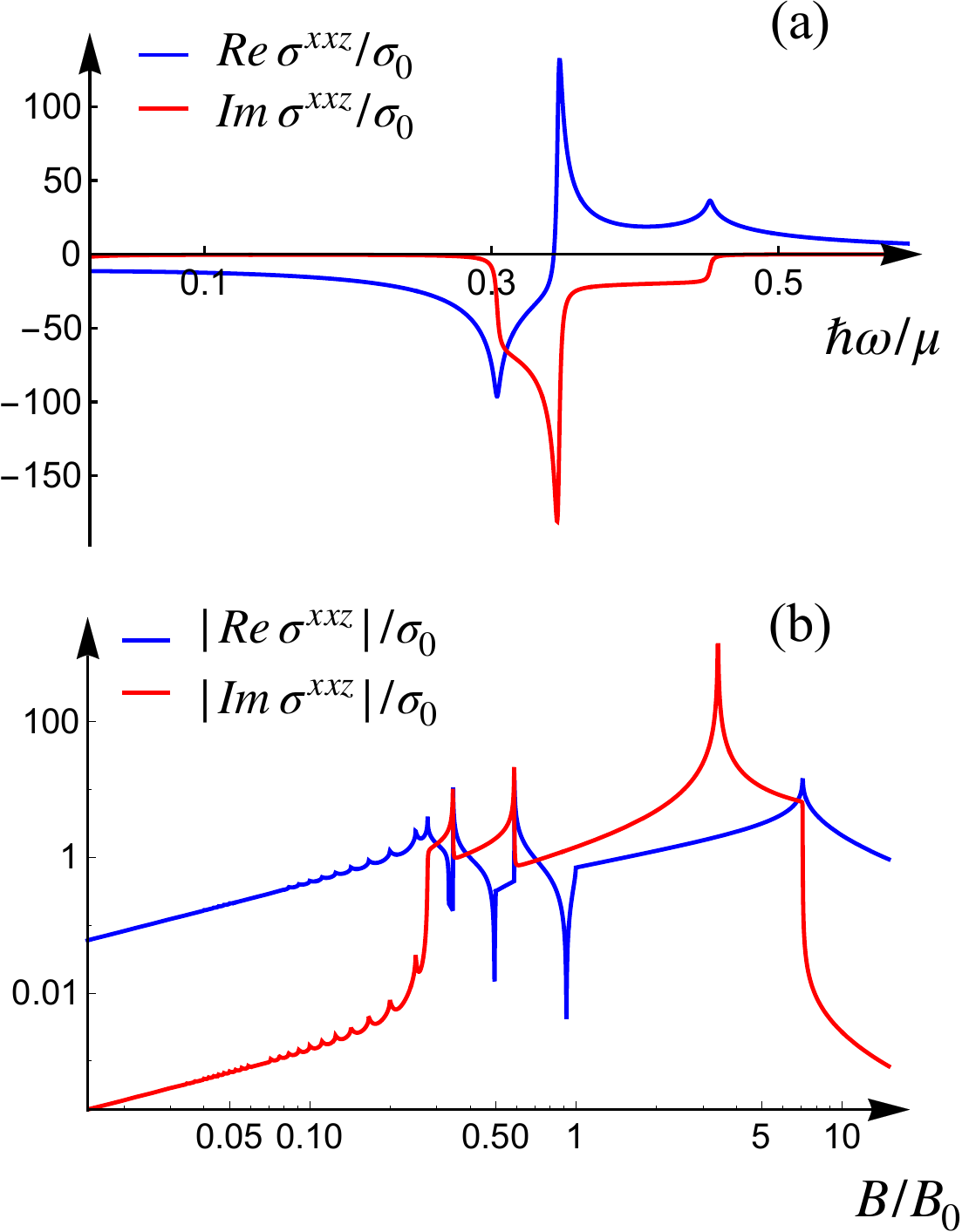}
    \caption{Real (blue) and imaginary (red) parts of the SHG component $\sigma^{xxz}$.  A small but finite imaginary part of frequency, which corresponds to a finite scattering rate, is kept to regularize the divergences. (a) The dependence on frequency at $B=0.7 B_0$, which demonstrates square-root and logarithmic divergences. Similar divergences but with smaller amplitude appear at higher frequencies and not shown in the figure. (b) The dependence on magnetic field at $\hbar \omega = 1.85 \mu$. The double logarithmic plot is used in order to catch different scales, so the absolute values of the conductivity components are shown. Quantum oscillations at small fields develop in a series of power-law and logarithmic divergences at higher fields. The units are $\sigma_0 \equiv \eta e^3/32 \pi^2 \hbar \mu$ and $B_0 \equiv \mu^2/2e\hbar v_F^2$.}
  \label{Fig:2nd1}
\end{figure}

Now we discuss the divergences of SHG components. The first set of logarithmic divergences in $\text{Re}\,\sigma^{xxz}$  occurs at 
\be
\hbar |\tilde\omega_1^{\pm}| = \sqrt{\hbar^2 \omega_B^2 + \mu^2}\pm \mu \label{Eq:omega_1}
\ee
discussed earlier in the context of the EQL. Furthermore, away from the quantum limit, at $\hbar \omega_B < |\mu|$, it has additional set of logarithmic divergences at 
\be \label{Eq:omega_2}
\hbar |\tilde \omega_2^{\pm}| =  |\mu| \pm \sqrt{\mu^2 - \hbar^2 \omega_B^2}.
\ee 
These resonant frequencies  exactly correspond to the boundaries of the frequency range defined by the inequalities in Eqs.~\eqref{Eq:region1} and~\eqref{Eq:region2} or~\eqref{Ineq:1}--\eqref{Ineq:3}, where the injection current is nonzero. They equal the energy differences between the adjacent Landau levels (or their opposite energy copies from a different band) at wave vectors $k_z = k_n/v_F$ where they cross the chemical potential, see Fig.~\ref{LandauLevelsPicture} and Eq.~\eqref{Eq:Nmax}. Finally, both $\text{Re}\,\sigma^{xxz}$ and $\text{Im}\,\sigma^{xxz}$ have one-sided power-law (square-root) divergences at 
\be 
|\tilde \omega_3^{\pm}| =  \omega_B \left( \sqrt{N_{\text{max}}+1} \pm \sqrt{N_{\text{max}}} \right), \label{Eq:omega_3} 
\ee
where $N_{\text{max}}$ is given by Eq.~\eqref{Eq:Nmax}. These are the resonant frequencies that equal the energy difference between the lowest empty  and the highest partially filled Landau levels (or its opposite energy copies) at momentum $k_z=0$. Formally, they are obtained from $\tilde \omega_1$ by the substitute $\mu^2 \to \hbar^2\omega_B^2 \left\lfloor\mu^2/(\hbar\omega_B)^2 \right\rfloor$. We note that, in the EQL, $\hbar \omega_B > |\mu|$,   $\text{Im}\,\sigma^{xxz}$ has stronger power-law divergence at $\tilde \omega_3$, while $\text{Re}\,\sigma^{xxz}$ remains finite at this frequency in the limit of vanishing scattering rate (the divergence reappears, however, when small but finite scattering rate is included). 

At low fields $\hbar \omega_B \ll |\mu|$, the resonant frequencies become 

\be \label{Eq:resapprox} 
\hbar\left|\tilde \omega_i^-\right| \approx \frac{\hbar^2 \omega_B^2}{2|\mu|}, \qquad \hbar\left|\tilde \omega_i^+\right| \approx 2|\mu|, \qquad i=1,2,3.
\ee 
Equations~\eqref{Eq:xxzloww} and~\eqref{Eq:resapprox} indicate that a giant SHG can be achieved even at low magnetic fields and low frequencies, provided the scattering rate is sufficiently small.
We discuss the asymptotic behavior in the vicinity of the divergences in more detail in Appendix~\ref{App:SHGxxz}.

\subsection{${\boldsymbol \sigma^{xzx}}$ component of SHG}
Next, we consider components $ \sigma^{xzx} = \sigma^{zxx} = \sigma^{zyy} = \sigma^{yzy}$. The general expression is derived in Appendix~\ref{App:SHGxzx} and given by

\begin{align}  
&\sigma^{xzx}(i\omega) \equiv \sigma^{xzx}(i\omega,i\omega) = 
\frac{\eta e^3}{32 \pi^2 \hbar^2}\frac{\omega_B^2}{\omega^2} \frac{\text{sgn}(\mu)}{4\omega^4 - \omega_B^4} \times \nonumber 
 \\ &\sum_{n=0}^{N_{\max}}\left[(\omega_B^2 + 2\omega^2)(\omega_B^2 + 4\omega^2)+12 \omega^2 \omega_B^2 n  \right]f_n(2 i\omega) - \nonumber \\ &\left[2(\omega_B^2 + \omega^2)(\omega_B^2 + 2\omega^2)+12 \omega^2 \omega_B^2 n  \right]f_n(i\omega), \label{Eq:SHGgeneralxzx}
\end{align}
where $f_n(i\omega)$ and $N_{\max}$ are defined in Eqs.~\eqref{Eq:fn}-\eqref{Eq:Nmax}. The dependence on frequency and magnetic field after analytic continuation is shown in Fig.~\ref{Fig:2ndxzx}, while the corresponding analytic expressions are presented in Appendix~\ref{App:SHGxzx}.

In the EQL, $\hbar \omega_B > |\mu|$, only the $n=0$ term contributes, and we find

\begin{align}
&\sigma^{xzx}(\omega) =\frac{\eta e^3 }{8 \pi^2 \hbar^2 }\frac{\omega_B^2 \, \text{sgn}(\mu)}{\omega(\omega_B^2+ 2\omega^2)} \times  \\ &\left\{\ln \left| \frac{2|\mu|\omega - \hbar(\omega_B^2-\omega^2)}{2|\mu|\omega + \hbar(\omega_B^2-\omega^2)} \cdot \frac{4|\mu|\omega + \hbar(\omega_B^2-4\omega^2)}{4|\mu|\omega - \hbar(\omega_B^2-4\omega^2)}  \right| + \pi i  \times \right. \nonumber \\  &\left. \left[ \Theta\left( 4|\omega\mu| - \hbar|\omega_B^2 - 4\omega^2|  \right)  -\Theta\left( 2|\omega\mu| - \hbar|\omega_B^2 - \omega^2|  \right)  \right]\right\}. \nonumber
\end{align}
The real part of this expression has logarithmical divergences at $\omega = \tilde \omega^{\pm}_1$ and $\omega = \tilde \omega^{\pm}_1/2$, where $\tilde \omega^{\pm}_1$ is defined in Eq.~\eqref{Eq:omega_1}, while its imaginary part is free of divergences.

\begin{figure}
  \centering
  \includegraphics[width=1.\columnwidth]{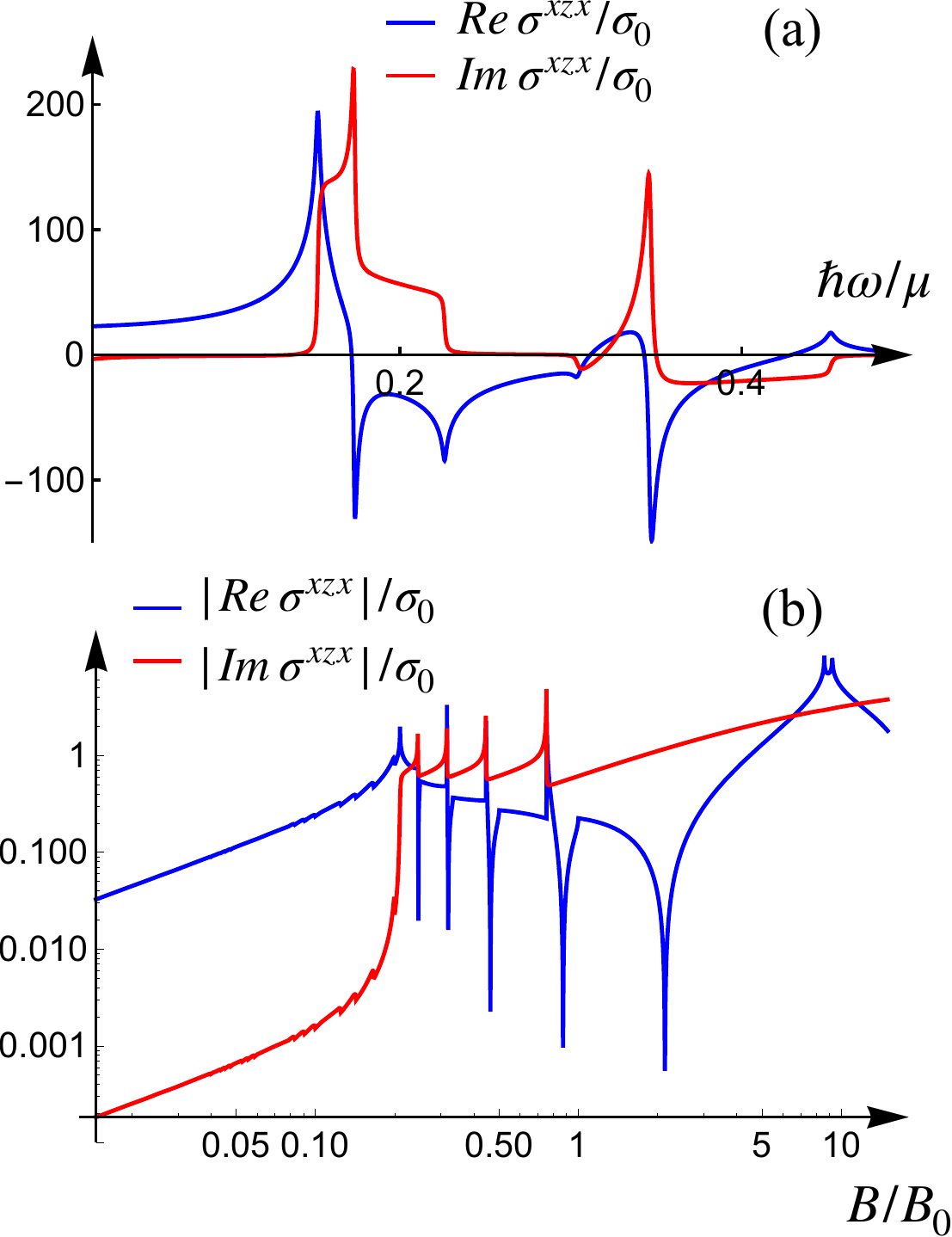}
    \caption{Real (blue) and imaginary (red) parts of the SHG component $\sigma^{xzx}$.  A small but finite imaginary part of frequency, which corresponds to a finite scattering rate, is kept to regularize the divergences. (a) The dependence on frequency at $B=0.7 B_0$. The main qualitative features are the same as for the $\sigma^{xzx}$ component, see Fig.~\ref{Fig:2nd1}.  (b) The dependence on magnetic field at $\hbar \omega = 2.1 \mu$. The double logarithmic plot is used in order to catch different scales. The units are $\sigma_0 \equiv \eta e^3/32 \pi^2 \hbar \mu$ and $B_0 \equiv \mu^2/2e\hbar v_F^2$.}
  \label{Fig:2ndxzx}
\end{figure}

At small magnetic fields $\hbar^2 \omega_B^2 \ll \hbar^2 \omega^2,$ $\mu^2$, $|\hbar^2 \omega^2 - 4\mu^2|$, we find

\begin{align} \label{Eq:sigmaxzxsmallB}
\sigma^{xzx}(\omega) &\approx - \frac{\eta e^3 \omega_B^2}{32 \pi^2 \hbar \omega^2} \left\{ \frac{4\mu}{4\mu^2 - \hbar^2\omega^2}    \right. \nonumber\\ &+ \left. \pi i \delta(\hbar|\omega| - 2|\mu|) \text{sgn}(\omega \mu) \right\}. 
\end{align}
At low frequencies, $\omega \to 0$, we obtain
\be  
\sigma^{xzx}(\omega) \approx  \frac{\eta e^3  \mu}{2 \pi^2 \hbar^3 \omega_B^2}. 
\ee 
Away from the EQL, at $\hbar \omega_B < |\mu|$, the divergences occur at $\omega = \tilde \omega^{\pm}_i$ and $\omega = \tilde \omega^{\pm}_i/2$, $i=1,2,3$, where all $\tilde \omega^{\pm}_i$ are defined in Eqs.~\eqref{Eq:omega_1}--\eqref{Eq:omega_3}. The character of  these divergences is the same as for the $\sigma^{xxz}$ component.

\subsection{${\boldsymbol \sigma^{xzy}}$ component of SHG}

Finally, we present the result for the components $\sigma^{xzy} = \sigma^{zxy} = -\sigma^{yzx} = -\sigma^{zyx}$. The general expression looks much more complicated compared to all the other components, so we find it convenient to subtract the contribution at zero chemical potential $\mu = 0$. This extra term cancels out upon summation over the nodes with the opposite chiralities, which is required to get the physically measurable result (we assume that these nodes have the same Fermi velocity $v_F$). We obtain
\begin{widetext}
\begin{align}
&\sigma^{xzy}(i\omega,\mu) \equiv \sigma^{xzy}(i\omega,i\omega,\mu) -\sigma^{xzy}(i\omega,i\omega,0) = \nonumber \\ &\frac{\eta e^3}{8\pi^2 \hbar^2} \cdot \frac{\omega_B^2}{\omega (\omega_B^2 - 2\omega^2)} \ln \left[ \frac{k_0 (\omega_B^2 + \omega^2) + k_1 (\omega_B^2 - \omega^2)}{k_0 (\omega_B^2 + \omega^2) - k_1 (\omega_B^2 - \omega^2)} \cdot  \frac{k_0 (\omega_B^2 + 4\omega^2) - k_1 (\omega_B^2 - 4\omega^2)}{k_0 (\omega_B^2 + 4\omega^2) + k_1 (\omega_B^2 - 4\omega^2)}  \cdot \frac{\hbar^2 a_0^2(i\omega) + \mu^2}{\hbar^2 a_0^2(2i\omega) + \mu^2} \cdot \frac{a_0^2(2i\omega)}{a_0^2(i\omega)}  \right] + \nonumber \\  &\frac{3\eta e^3}{16 \pi^2 \hbar^2}\frac{\omega_B^2}{\omega^2 (\omega_B^4 - 4\omega^4)} \sum_{n=1}^{N_{\max}}\left\{\frac{8n \omega_B^2 \omega^2}{\sqrt{4n\omega_B^2 + \omega^2}} \ln\frac{\sqrt{4n\omega_B^2 + \omega^2} k_0 - \omega k_n}{\sqrt{4n\omega_B^2 + \omega^2} k_0 + \omega k_n} + \frac{(\omega_B^2 + \omega^2)(\omega_B^2 + 2\omega^2) + 6 \omega^2 \omega_B^2 n}3 g_n(i\omega) - \right.     \nonumber \\ &\left. \frac{(\omega_B^2 + 2\omega^2)(\omega_B^2 + 4\omega^2) + 12 \omega^2 \omega_B^2 n}6 g_n(2i\omega)   \right\},  \quad \text{with}   \label{Eq:SHGgeneralxzy}
\end{align}
\be  \label{Eq:gn}
g_n(i\omega) = \frac1{a_n(i\omega)} \ln\left[ \frac{2\omega a_n(i\omega)k_0 + k_{n+1}(\omega_B^2 - \omega^2)}{2\omega a_n(i\omega)k_0 - k_{n+1}(\omega_B^2 - \omega^2)} \cdot \frac{2\omega a_n(i\omega)k_0 - k_{n}(\omega_B^2 + \omega^2)}{{2\omega a_n(i\omega)k_0 + k_{n}(\omega_B^2 + \omega^2)}} \right],
\ee 
while $a_n(i\omega)$, $k_n$, and $N_{\max}$ are defined in Eq.~\eqref{Eq:Nmax} (we note that $k_0 = |\mu|/\hbar$). The dependence on frequency and magnetic field after analytic continuation is shown in Fig.~\ref{Fig:2ndxzy}, while the corresponding analytic expressions and details of calculation are presented in Appendix~\ref{App:SHGxzy}.


In the EQL, $\hbar \omega_B > |\mu|$, we find after analytic continuation

\begin{align}\label{Eq:sigmaxzyEQL}
&\sigma^{xzy}(\omega) = -\frac{i\eta e^3}{8\pi^2 \hbar^2} \cdot \frac{\omega_B^2}{\omega \left(2\omega^2 + \omega_B^2\right)} \left\{ \ln  \left| \frac{\hbar^2\left(\omega_B^2-4 \omega^2\right)^2-16 \omega^2 \mu^2}{\hbar^2\left(\omega_B^2-\omega^2\right)^2-4 \omega^2 \mu^2} \cdot \frac{\left(\omega_B^2-\omega^2\right)^2}{\left(\omega_B^2-4\omega^2\right)^2}     \right| - \right.  \\  & \left. \pi i \, \text{sgn}(\omega)\left[ \Theta\left(4 | \omega \mu| - \hbar|\omega_B^2-4\omega^2|\right) \text{sgn}\left(\omega_B^2-4\omega^2\right) -  \Theta\left(2 | \omega \mu| - \hbar|\omega_B^2-\omega^2|\right) \text{sgn}\left(\omega_B^2-\omega^2\right)   \right]      \right\}. \nonumber
\end{align}

At small fields $\omega_B \to 0$, we use the Euler-Maclaurin formula to find
\begin{align}\label{Eq:sigmaxzysmallB}
&\sigma^{xzy}(\omega) \approx \frac{3i\eta e^3 \hbar}{8\pi^2} \cdot \frac{\omega_B^3 |\mu|}{\omega (\mu^2 - \hbar^2\omega^2)(4\mu^2 - \hbar^2\omega^2)} \zeta \left( -\frac12, \frac{\mu^2}{\hbar^2 \omega_B^2} - \left\lfloor  \frac{\mu^2}{\hbar^2 \omega_B^2} \right\rfloor   \right) -  \\ &\frac{i\eta e^3}{64\pi^2 \hbar^2} \cdot \frac{\omega_B^4}{\omega^5} \left\{ \frac{5\hbar^4 \omega^4-16 \hbar^2 \omega^2 \mu^2 + 48\mu^4}{(4\mu^2 - \hbar^2 \omega^2)^2} + \frac23 \ln \frac{2^6 \omega^6 \mu^6 |\mu^2 - \hbar^2 \omega^2|}{\omega_B^6(4\mu^2 - \hbar^2 \omega^2)^4}   \right\} - \nonumber \\ &\frac{\eta e^3}{16\pi} \cdot \frac{\omega_B^3}{\omega^2 |\mu|} \zeta \left( -\frac12, \frac{\mu^2}{\hbar^2 \omega_B^2} - \left\lfloor \frac{\mu^2}{\hbar^2 \omega_B^2} \right\rfloor  \right) \left\{\delta(\hbar|\omega| - |\mu|) -  \delta(\hbar|\omega| - 2|\mu|)   \right\}+ \nonumber  \\ &\frac{\eta e^3}{96\pi \hbar^2} \cdot \frac{\omega_B^4}{|\omega|^5} \left\{\Theta(\mu^2 - \hbar^2 \omega^2) -  4\Theta(4\mu^2 - \hbar^2 \omega^2) + 6 \mu^2 \delta'\left(\hbar |\omega| - 2|\mu|  \right) - 6 |\mu| \delta \left(\hbar |\omega| - 2|\mu|  \right)   \right\}, \nonumber
\end{align}
\end{widetext}
where $\zeta(s,x)$ is the Hurwitz zeta function, and $\lfloor x\rfloor$ is the floor function. The presence of the Hurwitz zeta function results in  pronounced quantum oscillations even at small fields. Unlike other components, this one is proportional to $\omega_B^3$ at the smallest fields. We also note that the term proportional to $\omega_B^4$ quickly takes over upon increasing field; hence, we keep it in the asymptotic expansion as well.

\begin{figure}
  \centering
  \includegraphics[width=1.\columnwidth]{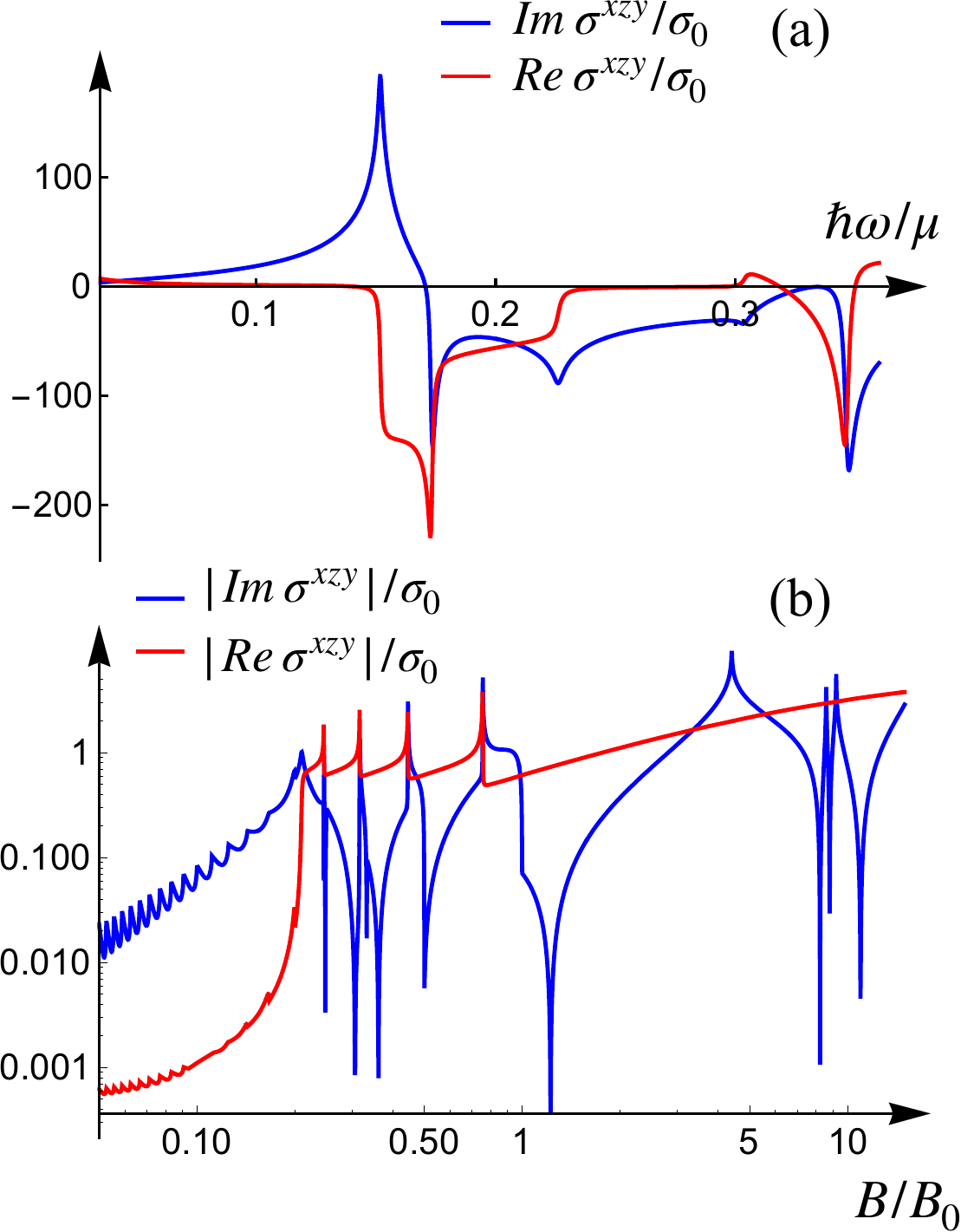}
    \caption{Real (red) and imaginary (blue) parts of the SHG component $\sigma^{xzy}$.  A small but finite imaginary part of frequency is kept to regularize the divergences. (a) The dependence on frequency at $B=0.7 B_0$. The main qualitative features are the same as for other components, see Figs.~\ref{Fig:2nd1}--\ref{Fig:2ndxzx}.  (b) The dependence on magnetic field at $\hbar \omega = 2.1 \mu$. The double logarithmic plot is used in order to catch different scales. The units are $\sigma_0 \equiv \eta e^3/32 \pi^2 \hbar \mu$ and $B_0 \equiv \mu^2/2e\hbar v_F^2$.}
  \label{Fig:2ndxzy}
\end{figure}

At low frequencies $\omega \to 0$, we obtain
\be \label{Eq:sigmaxzyloww}
\sigma^{xzy}(\omega) \approx \frac{3i\eta e^3}{4\pi^2\hbar^2} \cdot \frac{\omega}{\omega_B^2} \cdot \left( \frac{2\mu^2}{\omega_B^2 \hbar^2} + \frac{\hbar}{|\mu|}\sum_{n=1}^{N_{\max}}k_n \right),
\ee
which vanishes as frequency approaches 0.

Finally, we note that this component exhibits more divergences compared to all other components. Indeed, apart from the familiar divergences at  $\omega = \tilde \omega^{\pm}_i$ and $\omega = \tilde \omega^{\pm}_i/2$, $i=1,2,3$, defined in Eqs.~\eqref{Eq:omega_1}--\eqref{Eq:omega_3}, we also find logarithmical divergences at $|\omega| = \omega_B$, $\omega_B/2$, $2|\mu|/\hbar$ and power-law divergences at $|\omega| = 2\omega_B \sqrt{n}$, $\omega_B(\sqrt{n+1} + \sqrt{n})$, $\omega_B(\sqrt{n+1} + \sqrt{n})/2$, with $n=1,\ldots,N_{\max}$.

\section{Semiclassical description}
\label{Sec:semiclassics}

In this Section, we complement our microscopic result  with the semiclassical calculation. We neglect the interband and internodal transitions, focusing on a single band where the Fermi energy resides. This approach is justified provided $\hbar \omega \ll |\mu|$. Furthermore, we perform our analytic calculation to the linear order in small magnetic field $B$ only, which  requires additionally $\omega_B \ll \omega$. For large magnetic fields, a different approach is required, see, e.g., Ref.~\cite{BednikTikhonovSyzranov2020} for details.

The calculation scheme we implement is  similar to that of Refs.~\cite{Morimoto2016,Mandal2022}, though we point out certain differences in the results. The semiclassical equations of motion for an electron in a solid have the form~\cite{Xiaoreview2010}:
\begin{align}
\hbar \dot \br  &= \nabla_\bk \ve_\bk - \hbar \dot \bk \times {\bf \Omega}_{\bk}, \nonumber \\ 
\hbar \dot \bk &= -e \bE - e \dot\br \times \bB,
\end{align}
where 
\be
{\bf \Omega}_\bk = i \langle \nabla_{\bk} u_{\bk} | \times \nabla_{\bk} u_{\bk}    \rangle
\ee
is the Berry curvature and $|u_\bk\rangle$ is the periodic part of the Bloch wave function. The quasiparticle energy dispersion is modified according to $\ve_\bk = \ve^0_\bk - {\bf m}_\bk \cdot \bB$,  where $\ve^0_\bk$ is the bare band energy at $\bB=0$, $H_\bk |u_\bk\rangle = \ve^0_\bk |u_\bk\rangle$, and the orbital magnetic moment is given by 
\be 
{\bf m}_\bk = -i \frac{e}{2\hbar} \langle \nabla_{\bk} u_{\bk} | \times (H_\bk - \ve_\bk^0)| \nabla_{\bk} u_{\bk}    \rangle .
\ee

These equations can be readily resolved to give
\begin{align}
\dot \br &= \frac1{\hbar D_\bk} \left\{ \nabla_\bk \ve_\bk + e \bE \times {\bf \Omega}_\bk  + \frac{e}{\hbar} \bB \left( {\bf \Omega}_\bk \cdot \nabla_\bk \ve_\bk  \right) \right\}, \nonumber \\ \dot \bk &= \frac1{\hbar D_\bk}\left\{  -e \bE - \frac{e}{\hbar} \nabla_\bk \ve_\bk \times \bB - \frac{e^2}{\hbar} (\bE \cdot \bB) {\bf \Omega}_\bk  \right\}, \label{Eq:EOMresolved}
\end{align} 
where  $D_\bk = 1 + (e/\hbar) (\bB \cdot {\bf \Omega}_\bk)$ is the phase-space volume correction due to a finite Berry curvature~\cite{XiaoDOS2005}.

In the case of the {\it uniform} electric and magnetic fields, the kinetic equation for the distribution function $f$ in $\tau$-approximation is given by 
\be 
\frac{\partial f}{\partial t} + \dot \bk \nabla_\bk f = - \frac{f-f_0}{\tau},
\ee 
where we assume for simplicity that the scattering time $\tau$ is a constant. The zero-temperature equilibrium distribution function is given by $f_0=\Theta(\mu - \ve_\bk)$, where, again, $\Theta(x)$ is the Heaviside step function and $\mu$ is the chemical potential. The electrical current is then expressed as
\be 
{\bf j} = -e \int \frac{d^3k}{(2\pi)^3} D_\bk \dot \br f.
\ee 

A monochromatic electric field with frequency $\omega$ is given by 
\be
\bE(t) = \bE_\omega e^{-i\omega t} + \bE_{\omega}^* e^{i\omega t}, 
\ee
while the distribution function can be expanded in powers of small electric field, each corresponding to different harmonics: 
\be  
f(t) \approx f_0 + \delta f_0 + (f_1 e^{-i \omega t} +  f_1^* e^{i \omega t}) + (f_2 e^{-2i \omega t} + f_2^* e^{2i \omega t}).
\ee 
In the expression above, $f_1 \propto E$ and $f_2$, $\delta f_0 \propto E^2$, where $E = |\bE_\omega|$, and we only keep the terms up to order $\mathcal O(E^2)$.

The iterative solution of the kinetic equation in powers of $E$ is straightforward but rather lengthy and cumbersome.  Indeed, an algebraic mistake in solving the kinetic equation is what led to a disagreement in numerical coefficients for the semiclassical field-induced conductivity obtained in Sections III--V of Ref.~\cite{Morimoto2016}. We defer all the technical details of the calculation to Appendix~\ref{App:semiclassics}, and here, we only present the final result for the second-order conductivity components for a single Weyl node. In the vicinity of the node with the chirality $\eta = \pm1$, the Berry curvature and the orbital magnetic moment are given by the expressions
\be  
{\bf \Omega}_\bk = - \frac{\zeta \eta}{2 k^3} \bk, \qquad {\bf m}_\bk = - \eta \frac{e v_F}{2k^2} \bk, 
\ee 
while the band energy at zero field equals
\be
\ve^0_\bk = \zeta v_F \hbar k, 
\ee
where $\zeta = +1/-1$ corresponds to the conduction/valence band, respectively. 

Assuming the magnetic field $\bB = B \hat z$ is weak and keeping the terms linear in $B$ only (of the order of $\omega_B^2$), we find for the nonzero components of  dc conductivity

\begin{align} \label{Eq:semidc}
&\sigma^{xxz}_{\text{dc}} = \sigma^{yyz}_{\text{dc}} = - \frac{\eta e^3 \omega_B^2}{6 \pi^2 \hbar \mu} \cdot \frac{\tau^2}{(1+\omega^2\tau^2)^2},  \\ 
&\sigma^{xzx}_{\text{dc}} = \sigma^{yzy}_{\text{dc}} = \left( \sigma^{zxx}_{\text{dc}} \right)^* = \left( \sigma^{zyy}_{\text{dc}} \right)^* = \nonumber \\ &= \frac{\eta e^3 \omega_B^2}{24 \pi^2 \hbar \mu} \cdot \frac{\tau^2}{(1+\omega^2\tau^2)^2} \cdot \left[2+i\omega \tau \left(1 - \omega^2 \tau^2  \right)  \right], \nonumber 
\end{align}
and for SHG conductivity 
\begin{align} \label{Eq:semishg}
\sigma^{xzx}_{2\omega} &=  \sigma^{zxx}_{2\omega} =  \sigma^{yzy}_{2\omega} =  \sigma^{zyy}_{2\omega} = - \frac{\sigma^{xxz}_{2\omega}}2 = - \frac{\sigma^{yyz}_{2\omega}}2 = \nonumber \\ &= \frac{\eta e^3 \omega_B^2}{48 \pi^2 \hbar \mu} \cdot \frac{\tau^2(2-3i\omega\tau)}{(1-i\omega\tau)^2(1-2i\omega\tau)},
\end{align} 
while all other components equal 0. We emphasize that these semiclassical results are only valid provided $\omega_B \ll \omega \ll |\mu|/\hbar$.  

In the clean limit $\omega \tau \to \infty$, the nonvanishing components equal  
\begin{align} \label{Eq:semiclean}
\sigma^{xzx}_{2\omega} &=  \sigma^{zxx}_{2\omega} =  \sigma^{yzy}_{2\omega} =  \sigma^{zyy}_{2\omega} = - \frac{\sigma^{xxz}_{2\omega}}2 = - \frac{\sigma^{yyz}_{2\omega}}2 \approx \nonumber \\ &\approx - \frac{\eta e^3 \omega_B^2}{32 \pi^2 \hbar \omega^2 \mu}, \\ \sigma^{xzx}_{\text{dc}} &= \sigma^{yzy}_{\text{dc}} = \left( \sigma^{zxx}_{\text{dc}} \right)^* = \left( \sigma^{zyy}_{\text{dc}} \right)^* \approx \frac43i \sigma^{xzx}_{2\omega} \omega \tau. \nonumber
\end{align}
The SHG components are consistent with the low-field microscopic calculation in Sec.~\ref{Sec:SHG}, in particular, with Eqs.~\eqref{Eq:SHGsmallB} and~\eqref{Eq:sigmaxzxsmallB} at $\hbar\omega \ll |\mu|$. The nonzero components of the dc conductivity proportional to $\tau$, on the other hand, seem at first to contradict our Kubo formula results. Indeed, these components imply nonzero injection currents in the $x-$ and $y-$directions, while the calculation in Sec.~\ref{Sec:ShiftCurrent} indicates nonzero current in the $z-$direction only. The resolution of this seeming contradiction is hidden in the different order of noncommuting limits used in the two approaches. The microscopic calculation of Sec.~\ref{Sec:ShiftCurrent} implies taking the limit $\Omega \to 0$ first (equivalently, $\tau \to \infty$), before setting $\omega_B \to 0$. On the contrary, the semiclassical calculation of the present section assumes taking $\omega_B \to 0$ first, which leads to Eqs.~\eqref{Eq:semidc}--\eqref{Eq:semishg}, and only after that, the additional condition $\tau \to \infty$ results in Eq.~\eqref{Eq:semiclean}. While the order of limits is not important for the SHG components, it is crucial for calculating injection current, indicating that the latter depends sensitively on the parameter $\omega_B \tau$. This can be seen directly from the solution of the kinetic equation presented in detail in Appendix~\ref{App:semiclassics}. Similar seeming inconsistency originating from noncommuting limits was noticed when reproducing the fully isotropic $B=0$ result, see discussion in Sec.~\ref{Sec:ShiftCurrent} and Appendix~\ref{App:B=0}.


\section{Discussion}
\label{Sec:Discussion}

In summary, we studied the second-order optical response of a model with an ideal Weyl node in the presence of a magnetic field. Applying the Kubo formula, we calculated the intrinsic contribution to the conductivity components of photocurrent and second harmonic generation at arbitrary frequency, chemical potential, and magnetic field. Our results showed that the magnetic field significantly enhances the nonlinear response in general, while tuning the frequency (or magnetic field) to certain resonant values results in an unsaturated growth in magnitude. Although this growth would be eventually saturated in realistic systems due to disorder and interactions, our study suggests a method to realize photocurrents of extremely large magnitude.

Our results can be applied to Weyl materials with low crystal symmetry, such as SrSi$_2$~\cite{HasanSrSi22016}, RhSi~\cite{Reeseaba0509}, and TaAs~\cite{HasanTaAs2015,Wu2017} under strain, as well as other recently predicted topological semimetals~\cite{Hasantop2018,Cano2021}. While our exact analytical expressions were derived for type-I Weyl semimetals with linear dispersion, we expect qualitatively similar results (including divergences at resonant frequencies and magnetic fields) for other types of Weyl materials and topological band touching points. 

The nodes with different chiralities contribute oppositely to photocurrent and SHG, so they must be separated in energy in order to produce nonzero net current. This is only possible if inversion and all mirror symmetries are broken~\cite{deJuan2017}. This conclusion follows directly from the transformation properties of the Berry curvature which imply that the spatial symmetries relate the nodes with the opposite chiralities. The same logic dictates that time reversal symmetry can be preserved, since it relates the nodes with the same chirality. 

We emphasize again that the physically measurable response is obtained after summing up our analytical expressions over all the nodes with different chiralities. Since these nodes are located at different energies, they effectively experience ``different'' chemical potentials (with respect to the node location), thus leading to nonzero second-order current. The resulting expressions depend sensitively on the energy separation between the nodes and the position of the actual chemical potential, which itself depends on the magnetic field. We can comment, however, on two limiting cases which admit simple analytical interpretation. The first case is the nearly compensated semimetal with identical nodes (i.e., having the same velocity), such that the chemical potential is located almost exactly in between the nodes. In this case, the effective chemical potentials experienced by different nodes will have the same magnitude but opposite signs. This implies that the components with the even dependence on $\mu$ (i.e., CPGE component $\beta^{xyz}$, SHG component $\sigma^{xzy}$, and those related by symmetry) will cancel out upon summation, while the components having odd dependence on $\mu$  (i.e., LPGE component $\beta^{xxz}$, SHG components $\sigma^{xxz}$ and $\sigma^{xzx}$, and those related by symmetry) will enhance each other, merely multiplying our analytical expressions for a single node by the total number of the nodes. The second limiting case which can be easily analyzed is if the chemical potential nearly coincides with the position of one of the nodes,  i.e., its distance to the node is much smaller compared to other energy scales dictated by frequency and magnetic field. In this case, we can set $\mu \approx 0$ for one of the nodes, which immediately nullifies its contribution to all the components except for the CPGE component $\beta^{xyz}$. It means that  only nodes located away from chemical potential will contribute to the final result,  except for $\beta^{xyz}$ and other components related by symmetry.

Naturally, the overall second-order response in actual crystals may differ substantially from our analytical predictions derived for an idealized model~\cite{Wu2017,Haibin2018}. The main sources of discrepancy (apart from disorder and interactions) are possible tilt of the nodes, nonlinear corrections to the band dispersion, and the presence of higher-energy bands. We hope, however, that our results may guide experimental efforts to maximize the nonlinear response in topological materials.

As a potential future direction, we find it interesting to explore second-order responses in Weyl semimetals in the presence of pseudomagnetic fields originating from, e.g., external strain~\cite{Jorn2016,Pikulin2020,Sukhachov2020,Sukhachov2021}. The key difference from the physical magnetic field is that its coupling to Weyl electrons is chiral, which implies a substantially different result after summation over the nodes with the opposite chiralities. We leave a detailed analysis of this effect for future work.

Finally, we note that the method of Green's functions used in this paper can be easily extended to study the effects of interactions or disorder, as well as applied to other types of nodal points. In the absence of a magnetic field, the effect of Coulomb and short-ranged interactions on photocurrent was studied in Ref.~\cite{Avdoshkin2020}, while the role of impurities was investigated by the means of the kinetic equation in, e.g., Ref.~\cite{Levchenko2017}. The interplay between these factors and a magnetic field and, in particular, their effect on singularities discussed in this paper are open questions which we also leave for future study.


\section{Acknowledgements} 
We are grateful to Joel Moore, Justin Song, Takahiro Morimoto, Mingda Li, and especially Grigory Tarnopolsky for valuable discussions related to this project.

\appendix

\onecolumngrid

\section{Landau levels in a Weyl semimetal and matrix elements}
\label{App:LL}

The Hamiltonian for a Weyl fermion in magnetic field along the $z$-direction in Landau gauge $\bA = (- y B,0,0)$ is given by 

\be  
\hat H = \eta v_F \hbar \left[ \left( k_x - \frac{eB}{\hbar} y \right) \sigma_x -i\frac{\partial}{\partial y} \sigma_y + k_z \sigma_z \right],
\ee 
where $v_F$ is the Fermi velocity, $\eta = \pm 1$ is the chirality of the node, $\boldsymbol \sigma$ is the vector of Pauli matrices, $-e$ is the electron charge, and we assume $e>0$, $B>0$.

This Hamiltonian can be easily reduced to the harmonic oscillator problem. The eigenstates are

\begin{align}
&\Psi_{n,k_x,k_z}(\br) = e^{i k_x x + i k_z z} \sqrt{\frac1{2 L_x L_z}}\left( \begin{array}{c}   \sqrt{1 - \frac{E_0}{E_n}} \cdot f_{|n|-1}\left\{ \sqrt{\frac{eB}{\hbar}}\left[ y - y_0(k_x)\right] \right\}  \\  -\text{sgn}(n)\sqrt{1 + \frac{E_0}{E_n}} \cdot f_{|n|}\left\{ \sqrt{\frac{eB}{\hbar}}\left[ y - y_0(k_x)\right] \right\} \end{array} \right), \qquad n \ne 0, \nonumber \\ &\Psi_{0,k_x,k_z}(\br) = e^{i k_x x + i k_z z} \sqrt{\frac1{L_x L_z}} \left( \begin{array}{c}   0  \\ -f_{0}\left\{ \sqrt{\frac{eB}{\hbar}}\left[ y - y_0(k_x)\right] \right\}\end{array} \right),
\end{align}
where 
\begin{align}
&E_n = \text{sgn}(n) \eta \hbar v_F \sqrt{k_z^2 + \frac{2 e B}{\hbar} |n|}, \qquad n \ne 0, \nonumber \\ &E_0 = -\eta \hbar v_F k_z
\end{align}
is the energy spectrum, $y_0(k_x) = \hbar k_x/eB$ determines the center (``position'') of the wave function with momentum $k_x$, and $L_x$ and $L_z$ are linear dimensions of the system in the $x-$ and $z-$directions, correspondingly. We have introduced properly normalized harmonic oscillator eigenfunctions:

\be  
f_n(\xi) = \left(  \frac{eB}{\pi \hbar }\right)^{1/4} \frac1{\sqrt{2^n n!}} \exp\left( - \frac{\xi^2}{2} \right) H_n(\xi), \qquad n=0, 1, 2,\ldots
\ee 
with the Hermite polynomials

\be  
H_n(\xi) \equiv (-1)^n e^{\xi^2}\frac{d^n e^{-\xi^2}}{d \xi^n}.
\ee 

Using orthogonality of the Hermite polynomials, we find for the matrix elements

\begin{align}
    &\langle \Psi_n | \sigma_x | \Psi_m \rangle =   -\text{sgn}(m) \delta_{|m|,|n|-1} \sqrt{\frac{E_n - E_0}{2E_n}} \sqrt{\frac{E_m + E_0}{2E_m}} - \text{sgn}(n) \delta_{|n|,|m|-1} \sqrt{\frac{E_n + E_0}{2E_n}} \sqrt{\frac{E_m - E_0}{2E_m}}, \nonumber \\  &\langle \Psi_n | \sigma_y | \Psi_m \rangle =  i \, \text{sgn}(m) \delta_{|m|,|n|-1} \sqrt{\frac{E_n - E_0}{2E_n}} \sqrt{\frac{E_m + E_0}{2E_m}} -i\, \text{sgn}(n) \delta_{|n|,|m|-1} \sqrt{\frac{E_n + E_0}{2E_n}} \sqrt{\frac{E_m - E_0}{2E_m}}, \nonumber \\  &\langle \Psi_n | \sigma_z | \Psi_m \rangle = -\frac{E_0}{E_n} \delta_{n,m} + \sqrt{\frac{E_n^2 - E_0^2}{E_n^2}} \delta_{n,-m},  \label{SMEq:matrixelements}
\end{align}
where the wave functions $\Psi_m$ and $\Psi_n$ have the same quantum numbers $k_x$ and $k_z$, and we adopted the convention with $\text{sgn}(0)=1$.

\section{Second-order response in the limit $B \to 0$} \label{App:B=0}
In this Appendix, we reproduce the most general result for the second-order (three-current) correlation function for a Weyl node at $B=0$, obtained in Ref.~\cite{Avdoshkin2020}, from Eq.~\eqref{2ndOrderGeneric}. It is straightforward to check that, in the limit $B\to0$, only terms proportional to $\ve^{\alpha \beta \gamma}$ are nonzero, where $\ve^{\alpha \beta \gamma}$ is the fully antisymmetric Levi-Civita tensor. For concreteness, we consider the $\chi^{x y z}$ component. First, we find that 

\begin{align}
    Z_{n_1 n_2 n_3}^{xyz} = \frac{i}{4E_{n_1} E_{n_2}^2}\left\{\delta_{n_2,n_3} E_0\left[\delta_{|n_2|,|n_1|-1}  (E_{n_1}-E_0)(E_{n_2}+E_0) - \delta_{|n_2|,|n_1|+1} (E_{n_1}+E_0)(E_{n_2}-E_0)\right] \right. \nonumber \\ \left. + \delta_{n_2,-n_3}(E_{n_2}^2 - E_0^2) \left[ \delta_{|n_2|,|n_1|-1}  (E_{n_1}-E_0) + \delta_{|n_2|,|n_1|+1}  (E_{n_1}+E_0)  \right] \right\},
\end{align}
where we dropped argument $k_z$ in $E_{n_i}(k_z)$ for brevity. In the limit $B \to 0$, one may write $\delta_{|n_2|,|n_1|\pm1} \approx \delta_{|n_2|,|n_1|}$, and we obtain
\be  
Z_{n_1 n_2 n_3}^{xyz} \approx \frac{i \delta_{|n_1|,|n_2|}}{2 E_{n_1} E_{n_2}^2}\left\{\delta_{n_2, n_3}E_0^2 \left(E_{n_1} - E_{n_2}\right) + \delta_{n_2,-n_3}E_{n_1}\left(E_{n_2}^2 - E_0^2 \right)  \right\}.
\ee 
Next, we plug this expression into Eq.~\eqref{2ndOrderGeneric}. Utilizing delta-symbols and ignoring the contribution of the $n=0$ Landau level (which is negligible in the limit $B\to 0$), we obtain
\begin{align}
    \chi^{xyz}(i \omega_1, i\omega_2) &\approx \frac{i \eta e^4 v_F^3 B}{4\pi^2 \hbar} \sum_{n=-\infty}^{\infty} \int_0^{\infty} d k_z \frac{\Theta(\ve_n) - \Theta(\ve_{-n})}{E_n^2} \left\{ \frac{2 E_0^2}{(i\hbar \omega_1 - 2 E_n)(i\hbar \omega_2 + 2 E_n)} - \frac{E_n^2 - E_0^2}{(i\hbar \omega_1 + 2 E_n)(i\hbar \Omega + 2 E_n)}   \right.  \nonumber \\  &\left.+ \frac{E_n^2 - E_0^2}{(i\hbar \omega_2 + 2 E_n)(i\hbar \Omega + 2 E_n)} \right\} \approx - \frac{\eta e^4 v_F^3 B (\omega_1 - \omega_2)}{\pi^2} \sum_{n=1}^{\infty}\int_0^{\infty} d k_z \frac{\Theta(\ve_n) - \Theta(\ve_{-n})}{E_n}\times \\ &\times \frac{E_n^2\left[4 E_n^2 - \hbar^2 \omega_1 \Omega - \hbar^2 (\Omega + \omega_1)\omega_2\right] + E_0^2 \left[-12 E_n^2 - 2\hbar^2 \Omega^2 + \hbar^2 \Omega \omega_1 + \hbar^2 (\Omega + \omega_1) \omega_2\right]}{\left(4 E_n^2 + \hbar^2 \Omega^2\right)\left(4 E_n^2 + \hbar^2 \omega_1^2\right)\left(4 E_n^2 + \hbar^2 \omega_2^2\right)}, \nonumber
\end{align}
where we defined $\ve_n \equiv E_n(k_z) - \mu$, $\ve_{-n} \equiv E_{-n}(k_z) - \mu = -E_{n}(k_z) - \mu$ (since we excluded $n=0$), and $\Omega \equiv \omega_1 + \omega_2$. 

The limit $B\to0$ allows us to change the summation over $n$ with an integral. Furthermore, we find it convenient to introduce new variables $\rho \in (0, \infty)$ and $\varphi \in (0,\pi/2)$, such that

\be
\begin{array}{ccc} \hbar v_F k_z & = & \rho \cos \varphi \\ \hbar \omega_B \sqrt{n} & = & \rho \sin \varphi  \end{array} \qquad \Rightarrow  \qquad \begin{array}{ccc} E_{n\ne 0} & = & \eta \rho \\ |E_0| & = & \rho \cos \varphi \end{array} \qquad \text{and} \qquad d k_z d n = \frac{\rho^2 \sin\varphi d \rho d \varphi}{\hbar^2 e v_F^3 B}.
 \ee
Using these variables, the integral can be easily calculated:

\begin{align}
    &\chi^{xyz}(i \omega_1, i\omega_2) = -\frac{\eta e^3 (\omega_1 - \omega_2)}{\pi^2 \hbar^2 } \times \\&\int_{|\mu|}^{\infty}\rho^3 d \rho \int_0^{\pi/2} d\varphi \sin \varphi \frac{4 \rho^2 - \hbar^2 \omega_1 \Omega - \hbar^2 (\Omega + \omega_1)\omega_2 + \cos^2\varphi \left(-12 \rho^2 - 2\hbar^2 \Omega^2 + \hbar^2 \Omega \omega_1 + \hbar^2 (\Omega + \omega_1) \omega_2\right)}{\left(4 \rho^2 + \hbar^2 \Omega^2\right)\left(4 \rho^2 + \hbar^2 \omega_1^2\right)\left(4 \rho^2 + \hbar^2 \omega_2^2\right)}  \nonumber \\ &= \frac{\eta e^3}{48 \pi^2 \hbar^2} \cdot \frac{\Omega^3 (\omega_1 - \omega_2) \ln \left( 4\mu^2 + \hbar^2 \Omega^2 \right) + \omega_2^3(\Omega+\omega_1)\ln\left(4\mu^2 + \hbar^2 \omega_2^2\right) - \omega_1^3(\Omega+\omega_2)\ln\left(4\mu^2 + \hbar^2 \omega_1^2\right)}{\omega_1 \omega_2 \Omega}. \nonumber
\end{align}

All other nonzero components of $\chi^{\alpha \beta \gamma}$ can be calculated absolutely analogously. Collecting them all together, we find that, in the limit $B \to 0$, the correlation function is given by

\be  \label{SMEq:chi_B=0} 
\chi^{\alpha \beta \gamma}(i \omega_1, i\omega_2) = \frac{\eta e^3 \ve^{\alpha \beta \gamma}}{48 \pi^2 \hbar^2} \cdot \frac{\Omega^3 (\omega_1 - \omega_2) \ln \left( 4\mu^2 + \hbar^2 \Omega^2 \right) + \omega_2^3(\Omega+\omega_1)\ln\left(4\mu^2 + \hbar^2 \omega_2^2\right) - \omega_1^3(\Omega+\omega_2)\ln\left(4\mu^2 + \hbar^2 \omega_1^2\right)}{\omega_1 \omega_2 \Omega},
\ee
in full agreement with Ref.~\cite{Avdoshkin2020}.

\section{Injection current}
\label{App:injection}

The starting point for calculating the injection current is Eqs.~\eqref{Eq:injcurrent}--\eqref{Eq:betagen}. The form of the matrix elements entering this expression $Z_{n_1 n_2 n_2}^{\alpha \beta \gamma}$ and $Z_{n_1 n_2 n_2}^{\beta \alpha \gamma}$ dictates that only components with $\gamma = z$ are possibly nonzero, as follows from Eqs.~\eqref{Eq:Zs} and~\eqref{SMEq:matrixelements}. Indeed, we find that the only nonvanishing components of the injection current are $\beta^{xyz} = -\beta^{yxz}$ and $\beta^{xxz} = \beta^{yyz}$, and we discuss them in more detail below.

\subsection{CPGE component ${\boldsymbol \beta^{xyz}}$}

After substituting Eq.~\eqref{SMEq:matrixelements} into Eq.~\eqref{Eq:Zs}, we find that
\begin{align}
&Z^{xyz}_{n_1 n_2 n_2} = -Z^{yxz}_{n_1 n_2 n_2} = \frac{i E_0}{4 E_{n_1} E^2_{n_2}}\left\{\delta_{|n_2|,|n_1|-1}(E_{n_1} - E_0)(E_{n_2}+E_0)  - \delta_{|n_2|,|n_1|+1} (E_{n_1}+E_0)(E_{n_2}-E_0)  \right\}.
\end{align}
Plugging this into Eq.~\eqref{Eq:injcurrent}, we obtain: 

\begin{align}  
\sigma^{xyz}(\omega + \Omega, -\omega) = \frac{\eta e^4 v_F^3 B}{8\pi \hbar^2 \Omega \omega^2} \int_{-\infty}^{\infty} d k_z \sum_{n_1, n_2 = -\infty}^{\infty} \left[\Theta(\ve_1) - \Theta(\ve_2)  \right] \delta(\ve_2 - \ve_1 - \hbar \omega) \frac{E_0 (E_{n_1} - E_{n_2})}{E_{n_1}^2 E_{n_2}^2} \times \nonumber \\ \left[\delta_{|n_1|, |n_2|-1}\left(E_{n_2} - E_0\right)\left(E_{n_1} + E_0\right) - \delta_{|n_1|, |n_2|+1}\left(E_{n_1} - E_0\right)\left(E_{n_2} + E_0\right)  \right],
\end{align}
where, again, we defined $\ve_{1,2} \equiv E_{n_{1,2}}(k_z) - \mu$. It is straightforward to check that this expression is even with respect to chemical potential $\mu$ and odd with respect to frequency $\omega$. 

\begin{figure}
\centering
\includegraphics[width=1.\columnwidth]{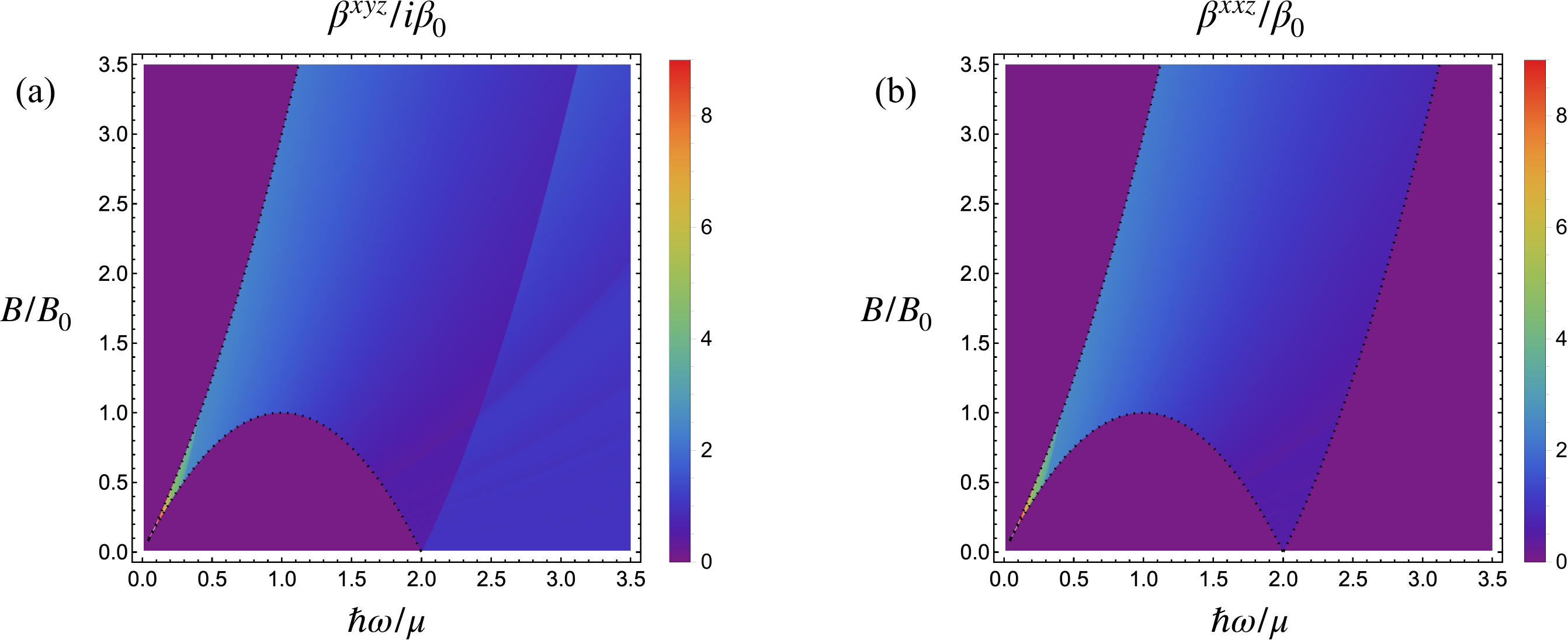}
\caption{Density plot of the CPGE component $\beta^{xyz}$ (a) and LPGE component $\beta^{xxz}$ (b), given by Eqs.~\eqref{Eq:betaxyz}--\eqref{Eq:betaxxz}, as a function of frequency and magnetic field. The black dotted lines indicate the boundaries of the regions where the corresponding components are nonzero, see Eqs.~\eqref{Eq:region1} and~\eqref{Eq:region2}. A characteristic Landau fan indicates fields at which higher Landau levels become partially populated. The units are $\beta_0 \equiv -\eta e^3/12 \pi \hbar^2$ and $B_0 \equiv \mu^2/2e\hbar v_F^2$.
}
\label{AppFig:densitypge}
\end{figure}

The Kronecker delta is used to perform summation over one of the Landau indices (positive, negative, and zero Landau levels should be considered separately), and the integration over $k_z$ is straightforward  due to the delta-function. Carefully collecting all the contributions, simplifying the resulting expression, and using Eq.~\eqref{Eq:betagen}, we arrive at Eq.~\eqref{Eq:betaxyz}. The density plot of this component as a function of frequency and magnetic field as well as the parameter range where it is nonzero are shown in Fig.~\ref{AppFig:densitypge}(a). We see that the photocurrent is enhanced around the line $B/B_0 = 2\hbar \omega/\mu$ at small frequencies, where $B_0 \equiv \mu^2/2 e \hbar v_F^2$, in agreement with Eqs.~\eqref{Eq:dcpeakpos}--\eqref{Eq:dcpeakrange}.

At vanishing field and fixed frequency and chemical potential $\omega_B \ll |\omega|, |\mu|/\hbar$, we may change the summation over Landau levels with integration and get 

\begin{align}
\beta^{xyz}(\omega) &\approx -\frac{i \eta e^3}{2\pi \hbar^2} \frac{\omega_B^2 \text{sgn} (\omega)}{\omega^4} \Theta\left(\hbar|\omega|  - 2 |\mu|\right)  \int_0^{\omega^2/4\omega_B^2} dn \, \sqrt{\omega^4 -4\omega^2 \omega_B^2 n} = -\frac{i \eta e^3}{12\pi \hbar^2}\text{sgn} (\omega)\Theta\left(\hbar|\omega|  - 2 |\mu|\right) = \nonumber \\ &= i \beta_0 \text{sgn} (\omega)\Theta\left(\hbar|\omega|  - 2 |\mu|\right), \qquad \text{with} \quad \beta_0 \equiv -\frac{\eta e^3}{12 \pi \hbar^2},
\end{align}
in full agreement with Refs.~\cite{deJuan2017,Avdoshkin2020},  as well as Eq.~\eqref{SMEq:chi_B=0} in the limit $\Omega \to 0$. We note, however, that the corresponding current only flows in the $z-$direction (along magnetic field),  unlike Eq.~\eqref{SMEq:chi_B=0}, since we took the limit $\Omega \to 0$ before setting $\omega_B \to 0$.

In the low-frequency and low-field limit $|\omega| \ll \omega_B \ll 2|\mu|/\hbar$, the system can be tuned to the resonance frequency $|\omega| \approx \hbar\omega_B^2/2|\mu|$, where the current is significantly enhanced. Indeed, in the regime 
\be  
\left|\hbar\omega_B^2 - 2\left|\omega \mu\right|\right| < \hbar\omega^2 \ll \hbar\omega_B^2 \approx 2\left|\omega \mu\right|,
\ee 
we find that $\Theta\left(\hbar(\omega^2 + \omega_B^2) - 2 |\omega \mu|\right) = 1$,  $\Theta\left(\hbar|\omega^2 - \omega_B^2| - 2|\omega \mu| \right) = 0$, while Eq.~\eqref{Eq:betaxyz} leads to

\be  
\beta^{xyz}(\omega) \approx -\frac{i \eta e^3}{4\pi \hbar^2} \frac{\text{sgn} (\omega)}{\omega_B^2} \int_0^{\omega_B^2/4\omega^2} dn \,\sqrt{\omega_B^4 -4\omega^2 \omega_B^2 n} \approx  - i \frac{\eta e^3}{24 \pi \hbar^2} \frac{\omega_B^2}{\omega^2} \text{sgn} (\omega) \approx i \beta_0 \frac{|\mu|}{\hbar\omega}, \qquad \left| \beta^{xyz}(\omega) \right| \gg |\beta_0|,
\ee
where, again, we changed the summation with the integral.

\subsection{LPGE component ${\boldsymbol \beta^{xxz}}$}

The whole procedure for calculating $\beta^{xxz} = \beta^{yyz}$ is analogous to that for CPGE components. The relevant matrix elements are now given by
\begin{align}
&Z^{xxz}_{n_1 n_2 n_2} = Z^{yyz}_{n_1 n_2 n_2} = -\frac{E_0}{4 E_{n_1} E^2_{n_2}}\left\{\delta_{|n_2|,|n_1|-1}(E_{n_1} - E_0)(E_{n_2}+E_0)  + \delta_{|n_2|,|n_1|+1} (E_{n_1}+E_0)(E_{n_2}-E_0)  \right\}.
\end{align}
In combination with Eq.~\eqref{Eq:injcurrent}, this gives
\begin{align}  
\sigma^{xxz}(\omega + \Omega, -\omega) = -i\frac{\eta e^4 v_F^3 B}{8\pi \hbar^2 \Omega \omega^2} \int_{-\infty}^{\infty} d k_z \sum_{n_1, n_2 = -\infty}^{\infty} \left[\Theta(\ve_1) - \Theta(\ve_2)  \right] \delta(\ve_2 - \ve_1 - \hbar \omega) \frac{E_0 (E_{n_1} - E_{n_2})}{E_{n_1}^2 E_{n_2}^2} \times \nonumber \\ \left[\delta_{|n_1|, |n_2|-1}\left(E_{n_2} - E_0\right)\left(E_{n_1} + E_0\right) + \delta_{|n_1|, |n_2|+1}\left(E_{n_1} - E_0\right)\left(E_{n_2} + E_0\right)  \right],
\end{align}
which after integration over $k_z$ and summation over Landau indices leads to Eq.~\eqref{Eq:betaxxz}. The density plot for this component is shown in Fig.~\ref{AppFig:densitypge}(b). The structure of step functions dictates that this component is nonzero only provided $\hbar(\omega^2 + \omega_B^2) - 2 |\omega \mu| > 0$ and $\hbar|\omega^2 - \omega_B^2| - 2|\omega \mu| < 0$. In this parameter range, which is shown in Fig.~\ref{AppFig:densitypge}(b) by the black dotted lines, different components are related according to
\be
\beta^{xxz}(\omega) = -i \beta^{xyz}(\omega)\text{sgn}(\omega \mu).
\ee
Finally, in the low-field limit, $\omega_B \ll |\omega|, |\mu|/\hbar$, the two step functions can be combined into a single one, and we obtain

\begin{align}
\beta^{xxz}(\omega) &\approx -\frac{\eta e^3}{4\pi \hbar^2} \frac{\omega_B^2 \text{sgn} (\mu)}{\omega^4} \Theta\left(\frac{\omega_B^2}{|\omega|}-\left||\omega| - 2\frac{|\mu|}{\hbar}  \right| \right)  \int_0^{\omega^2/4\omega_B^2} dn \, \sqrt{\omega^4 -4\omega^2 \omega_B^2 n} = \nonumber \\ &= \frac{\beta_0}2\text{sgn}(\mu)\Theta\left(\frac{\omega_B^2}{|\omega|}-\left||\omega| - 2\frac{|\mu|}{\hbar}  \right| \right) , \qquad \text{with} \quad \beta_0 \equiv -\frac{\eta e^3}{12 \pi \hbar^2}.
\end{align}
This expression is nonzero only in the narrow frequency range of the width $2\omega_B^2/|\omega| \approx \hbar \omega_B^2/|\mu|$ near $\hbar |\omega| = 2|\mu|$.

\section{Second harmonic generation}
\label{App:SHG}

For an ideal Weyl node, the only nonzero SHG components are $\sigma^{xxz} = \sigma^{yyz}$, $\sigma^{zxx} = \sigma^{xzx} = \sigma^{zyy} = \sigma^{yzy}$, and $\sigma^{xzy} = \sigma^{zxy} = -\sigma^{yzx} = -\sigma^{zyx}$. It is straightforward to see from the structure of $Z^{\alpha\beta\gamma}_{n_1 n_2 n_3}$ factors, Eq.~\eqref{Eq:Zs}, that they equal zero if the number of $z-$components is even. Furthermore, SHG components $\sigma^{xyz} = \sigma^{yxz}=0$ vanish due to the identity $Z^{xyz}_{n_1 n_2 n_3} + Z^{yxz}_{n_1 n_2 n_3} =0$, and the component $\sigma^{zzz} =0$ becomes zero after integration over $k_z$. Below, we consider the derivation and limiting cases for all the nonzero components in more detail.

\subsection{${\boldsymbol \sigma^{xxz}}$ component of SHG}
\label{App:SHGxxz}

We start with the simplest nonzero component $\sigma^{xxz}(i\omega) = \chi^{xxz}(i\omega, i\omega)/\omega^2$. We note that there is only one term in this expression, which merely accounts for the fact that both electric field components in this case have the same frequency and polarization and are physically indistinguishable. 

The starting point for the calculation is Eq.~\eqref{2ndOrderGeneric}. The corresponding form-factor equals

\begin{align}
Z^{xxz}_{n_1 n_2 n_3} &= -\frac{E_0}{4 E_{n_1} E^2_{n_2}} \delta_{n_2,n_3} \left[\delta_{|n_1|,|n_2|-1}(E_{n_1}+E_0)(E_{n_2}-E_0) + \delta_{|n_2|,|n_1|-1}(E_{n_2}+E_0)(E_{n_1}-E_0)  \right]  \nonumber \\ &-\frac{E_{n_2}^2-E_0^2}{4 E_{n_1} E^2_{n_2}} \delta_{n_2,-n_3} \left[\delta_{|n_2|,|n_1|-1}(E_{n_1}-E_0) - \delta_{|n_1|,|n_2|-1}(E_{n_1}+E_0)\right].
\end{align}
Plugging this expression into Eq.~\eqref{2ndOrderGeneric} and performing summation over one of the Landau indices, we obtain

\begin{align}
\sigma^{xxz}(i\omega) &= -\frac{\eta e^4 v_F^3 B}{16 \pi^2 \hbar \omega^2}\int_{-\infty}^{\infty} d k_z
 \sum_{n_1, n_2} \frac{\Theta(\ve_1) - \Theta(\ve_2)}{\hbar^2\omega^2 + (E_{n_1} - E_{n_2})^2} \cdot \frac{E_0}{E_{n_1} E_{n_2}^2} \times \nonumber \\ &\times \left[ \delta_{|n_1|, |n_2|-1} (E_{n_1} + E_0)(E_{n_2} - E_0) + \delta_{|n_2|, |n_1|-1} (E_{n_2} + E_0)(E_{n_1} - E_0) \right] + \nonumber \\ &+ \frac1{i\hbar\omega + E_{n_1} + E_{n_2}} \left[ \frac{\Theta(\ve_2) - \Theta(\ve_1)}{i\hbar\omega + E_{n_2} - E_{n_1}}-  \frac{\Theta(\ve_2) - \Theta(\ve_{-2})}{2i\hbar\omega + 2E_{n_2} }\right] \frac{ E_{n_2}^2 - E_0^2}{E_{n_1} E_{n_2}^2} \times \nonumber\\ &\times \left[ \delta_{|n_2|, |n_1|-1} (E_{n_1} - E_0) - \delta_{|n_1|, |n_2|-1} (E_{n_1} + E_0)\right],
\end{align}  
where $\ve_i = E_{n_i}(k_z) - \mu$ and $\ve_{-i} = -E_{n_i}(k_z) - \mu$. We find it convenient to further split this expression into the quantum contribution $\sigma^{xxz}_{\text{q}}$, involving $n=0$ Landau level, and the nonquantum part $\sigma^{xxz}_{\text{nq}}$, with $n_{1,2}\ne0$ only. After some simplification, the former is given by 

\be
\sigma^{xxz}_{\text{q}} = -\frac{\eta e^4 v_F^3 B}{16 \pi^2 \hbar \omega^2}\text{v.p.}\int_{-\infty}^{\infty} d k_z \frac{2}{E_0 E_1} \left\{\left[ \Theta(\ve_{-1}) - \Theta(\ve_0)  \right] \frac{(E_0 + E_1)^2}{\hbar^2 \omega^2 + (E_0 + E_1)^2}  -  \left[ \Theta(\ve_{1}) - \Theta(\ve_0)  \right] \frac{(E_0 - E_1)^2}{\hbar^2 \omega^2 + (E_0 - E_1)^2}\right\},
\ee 
where v.p. stands for the Cauchy principal value and $\ve_{\pm1} = \pm E_1(k_z) - \mu$ refers to the first Landau level in this particular equation, while the latter equals

\be
\sigma^{xxz}_{\text{nq}} = -\frac{\eta e^4 v_F^3 B}{16 \pi^2 \hbar \omega^2} \sum_{n_1, n_2 \ne 0} \int_{0}^{\infty} d k_z \frac{\Theta(\ve_{2}) - \Theta(\ve_1) }{E_{n_1} E_{n_2}} \frac{(E_{n_2} - E_{n_1})^2}{\hbar^2 \omega^2 + (E_{n_2} - E_{n_1})^2} \left( \delta_{|n_2|, |n_1|-1} - \delta_{|n_1|, |n_2|-1} \right).
\ee
It is straightforward to check that both quantum and nonquantum components are  odd functions of chemical potential $\mu$ and node chirality $\eta$.  Focusing on $\mu > 0$ and $\eta = +1$ for simplicity, one may perform the summation over one of the Landau indices explicitly (again, positive and negative Landau levels should be considered separately). Upon simplifying the resulting expressions, we obtain

\be  \label{AppEq:sigmaxxz}
\sigma^{xxz}(i\omega) = \sigma^{xxz}_{\text{q}} + \sigma^{xxz}_{\text{nq}} = - \frac{e^4 v_F^3 B}{2 \pi^2 \hbar^3} \sum_{n=0}^{\infty}\int_0^{\infty} d k_z \frac{\Theta(\ve_{n+1}) - \Theta(\ve_n)}{(\omega^2 + \omega_B^2)^2 + 4\omega^2 (\omega_B^2 n + v_F^2 k_z^2)},
\ee 
where $n=0$ term exactly corresponds to the quantum contribution. Integration over $k_z$ can now be performed explicitly leading to Eqs.~\eqref{Eq:SHGgeneralxxz}--\eqref{Eq:fn}, where we restored generic $\eta$ and $\mu$. We note that this expression is purely real and, in principle, could be rewritten through $\arctan$ function. However, we keep logarithms for the purpose of analytic continuation that we discuss next. 

The analytic continuation, $i\omega \to \omega + i0$, can be performed either from Eq.~\eqref{Eq:fn} or directly from integral in Eq.~\eqref{AppEq:sigmaxxz}. The result reads as 

\begin{align} \label{AppEq:sigmaxxzReIm}
\text{Re}\, \sigma^{xxz}(\omega) &= \frac{\eta e^3\omega_B^2 \,\text{sgn} (\mu)}{32 \pi^2 \hbar^2 \omega^2} \left\{  \sum_{n=0}^{N_{0}(\omega)}\frac{1}{|a_n(\omega)|}\ln \left|\frac{(k_n-|a_n(\omega)|)(k_{n+1}+|a_n(\omega)|)}{(k_n+|a_n(\omega)|)(k_{n+1}-|a_n(\omega)|)}  \right|  + \right. \nonumber \\ &\left. \sum_{n=N_0(\omega)+1}^{N_{\max}} \frac2{a_n(\omega)}\arctan \left[ \frac{a_n(\omega)(k_n-k_{n+1})}{k_n k_{n+1} + a_n^2(\omega)}  \right] \right\},  \\ \text{Im}\, \sigma^{xxz}(\omega) &= -\frac{\eta e^3\omega_B^2 \,\text{sgn} [\mu \omega (\omega_B^2 - \omega^2)]}{32 \pi \hbar^2 \omega^2}   \sum_{n=0}^{N_{0}(\omega)}\frac{\Theta[(k_n - |a_n(\omega)|)(|a_n(\omega)|- k_{n+1})]}{|a_n(\omega)|}, \nonumber
\end{align}
where we defined
\be  \label{AppEq:defs}
N_{\max} \equiv \left\lfloor\frac{\mu^2}{\hbar^2\omega_B^2}\right\rfloor, \quad N_0(\omega) = \min\left\{ \left\lfloor\frac{(\omega_B^2 - \omega^2)^2}{4\omega^2 \omega_B^2}\right\rfloor, N_{\max}\right\}, \quad k_n \equiv \text{Re}\sqrt{\frac{\mu^2}{\hbar^2} - \omega_B^2 n},  \quad a_n(\omega) \equiv \sqrt{\omega_B^2 n - \frac{(\omega_B^2 - \omega^2)^2}{4\omega^2}}, 
\ee 
in accordance with Eq.~\eqref{Eq:Nmax}. We note that $a_n(\omega)$ is purely imaginary for $0\le n \le N_0(\omega)$ (such that $|a_n(\omega)| = [(\omega_B^2 - \omega^2)^2/4\omega^2 - \omega_B^2 n]^{1/2}$) and purely real for $N_0(\omega) + 1 \le n \le N_{\max}$ (provided $N_{\max} > N_0(\omega)$). To derive this result, we used analytic continuation of functions $f_n(i\omega)$ from Eq.~\eqref{Eq:fn}:

\begin{align}  \label{AppEq:fpm}
f_n(i\omega)  \quad \Longrightarrow \quad &f_n^+(\omega) \equiv -\frac2{a_n(\omega)}\arctan \left[ \frac{a_n(\omega)(k_n-k_{n+1})}{k_n k_{n+1} + a_n^2(\omega)}  \right], \quad &&\text{for} \quad n > \frac{(\omega_B^2 - \omega^2)^2}{4 \omega_B^2 \omega^2}, \nonumber\\ f_n(i\omega)  \quad \Longrightarrow \quad &f_n^-(\omega) \equiv\frac{1}{|a_n(\omega)|}\ln \left|\frac{(k_n+|a_n(\omega)|)(k_{n+1}-|a_n(\omega)|)}{(k_n-|a_n(\omega)|)(k_{n+1}+|a_n(\omega)|)}  \right| +  \\ +&\frac{\pi i \,\text{sgn}[\omega(\omega_B^2 - \omega^2)]}{|a_n(\omega)|} \Theta[(k_n - |a_n(\omega)|)(|a_n(\omega)|- k_{n+1})],  \quad &&\text{for} \quad n < \frac{(\omega_B^2 - \omega^2)^2}{4 \omega_B^2 \omega^2}. \nonumber
\end{align}

\begin{figure}
\centering
\includegraphics[width=1.\columnwidth]{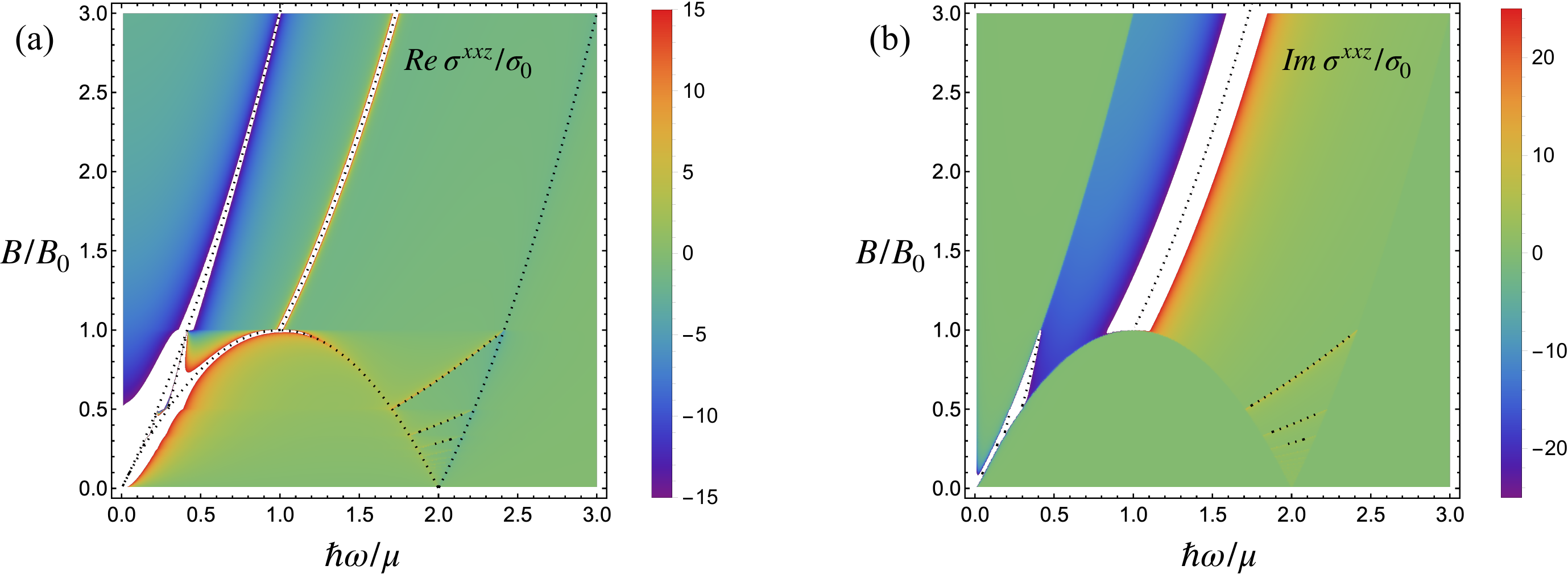}
\caption{Density plot of real (a) and imaginary (b) parts of $\sigma^{xxz}$ as a function of frequency and magnetic field, described by Eq.~\eqref{AppEq:sigmaxxzReIm}. We keep small but finite imaginary part of frequencies to mimic the effect of a finite scattering rate. Black dotted lines indicate the positions of the divergences given by Eqs.~\eqref{Eq:omega_1}--\eqref{Eq:omega_3}. White regions around the divergences imply that the conductivity components reach values (both positive and negative) larger than those  shown in the figure legend. The units are $\sigma_0 \equiv \eta e^3/32 \pi^2 \hbar \mu$ and $B_0 \equiv \mu^2/2e\hbar v_F^2$.
}
\label{AppFig:dplot2wxxz}
\end{figure}

We present the density plot of both real and imaginary parts of $\sigma^{xxz}(\omega)$ as a function of frequency and magnetic field in Fig.~\ref{AppFig:dplot2wxxz}. A small but finite imaginary part of frequency is added to mimic the effect of a finite scattering rate. It can be shown that the region with nonzero $\text{Im} \, \sigma^{xxz}$ exactly corresponds to the region with nonzero LPGE component $\beta^{xxz}$, which is given by Eq.~\eqref{Eq:region2}. The boundaries of this region are accompanied with some of the divergences of $\text{Re} \, \sigma^{xxz}$, Eqs.~\eqref{Eq:omega_1}--\eqref{Eq:omega_2}, which we discuss in more detail below. The line $B = B_0$ indicates the boundary of the extreme quantum limit, i.e., the first nonchiral Landau level starts populating below this field.

Equation~\eqref{AppEq:sigmaxxzReIm} is convenient for analyzing various limiting cases and divergences. In particular, we find that, at $\hbar \omega_B < |\mu|$ (away from the EQL), divergences take place at frequencies given by Eqs.~\eqref{Eq:omega_1}--\eqref{Eq:omega_3}, while corresponding asymptotic expressions in their vicinity take the form (we focus on $\omega, \, \mu >0$ for simplicity here):

\begin{align}
&\sigma^{xxz}(\omega) \approx \frac{\eta e^3}{32\pi^2 \hbar^2} \cdot \left(\frac{\omega_B}{\tilde\omega_1^{\pm}}\right)^2 \cdot \left( \sum_{n=0}^{N_{\max}}\frac1{k_n}  \right) \cdot \ln\left|\omega - \tilde\omega_1^{\pm}  \right| \quad &&\text{near} \quad \hbar |\tilde\omega_1^{\pm}| = \sqrt{\hbar^2 \omega_B^2 + \mu^2} \pm \mu,  \nonumber \\  &\sigma^{xxz}(\omega) \approx \frac{\eta e^3}{32\pi^2 \hbar^2} \cdot \left(\frac{\omega_B}{\tilde\omega_2^{\pm}}\right)^2 \cdot \left( \sum_{n=1}^{N_{\max}}\frac1{k_n}  \right) \cdot \ln\frac1{\left|\omega - \tilde\omega_2^{\pm}  \right|} \quad &&\text{near} \quad \hbar |\tilde \omega_2^{\pm}| = |\mu| \pm \sqrt{\mu^2 - \hbar^2 \omega_B^2}, \nonumber \\ &\sigma^{xxz}(\omega) \approx \frac{\eta e^3}{32\sqrt{2}\pi \hbar^2}\cdot \frac1{(\sqrt{N_{\max}+1}\pm \sqrt{N_{\max}})^{3/2}} \times  \\    &\times \left[ \frac1{N_{\max}(N_{\max}+1)}\right]^{1/4} \cdot \frac{\Theta\left[ \mp(\omega - \tilde \omega_3^{\pm}) \right] \pm i \Theta\left[ \pm(\omega - \tilde \omega_3^{\pm}) \right] }{\sqrt{\omega_B |\omega - \tilde  \omega_3^{\pm}|}} \quad &&\text{near} \quad   |\tilde \omega_3^{\pm}| =  \omega_B \left( \sqrt{N_{\text{max}}+1} \pm \sqrt{N_{\text{max}}} \right). \nonumber
\end{align}
These divergences are indicated by black dotted lines in Fig.~\ref{AppFig:dplot2wxxz}. The white regions around the divergences imply strong enhancement of $\sigma^{xxz}$ where it goes beyond the values shown in the plot legend. We see, in particular, that this component becomes very large even at small frequencies and magnetic fields, around the line $\hbar \omega/\mu = B/2B_0$ (equivalently, $\omega = \hbar  \omega_B^2/2\mu$), see Eq.~\eqref{Eq:resapprox}.

In the EQL, $\hbar \omega_B > |\mu|$, only the $n=0$ term contributes in Eq.~\eqref{AppEq:sigmaxxzReIm}, and we obtain Eq.~\eqref{Eq:EQLxxz}.

In the limit of vanishing magnetic field $\omega_B \to 0$, it is more convenient to start with the general expression in Matsubara frequencies given by Eqs.~\eqref{Eq:SHGgeneralxxz}--\eqref{Eq:Nmax}. In this limit, one may write
\be
N_{\max} \approx \frac{\mu^2}{\hbar^2 \omega_B^2}, \qquad a_n(i\omega) \approx \sqrt{\omega_B^2 n + \frac{\omega^2}4}, \qquad f_n(i\omega) \approx - \frac{4 \omega_B}{4\omega_B^2 N_{\max} + \omega^2} \cdot \frac1{\sqrt{N_{\max} -n}}.
\ee 
The summation over $n$ can then be replaced with integration, and we find 
\be  
\sigma^{xxz}(i\omega) \approx - \frac{\eta e^3 \omega_B^3 \, \text{sgn}(\mu)}{8\pi^2 \hbar^2 \omega^2 (4 \omega_B^2 N_{\max} + \omega^2)} \int_0^{N_{\max}} \frac{dn}{\sqrt{N_{\max}-n}} = - \frac{\eta e^3 \omega_B^2 \mu}{4\pi^2 \hbar \omega^2 (4 \mu^2 + \hbar^2\omega^2)}.
\ee
When performing analytic continuation, we use the identity
\be  
\frac1{4\mu^2 + \hbar^2 \omega^2} \quad \longrightarrow \quad \frac1{4\mu^2 + \hbar^2 (-i\omega + 0)^2} = \frac1{4\mu^2 - \hbar^2 \omega^2} + \frac{\pi i \, \text{sgn}(\omega \mu)}{4\mu} \delta(\hbar |\omega| - 2|\mu|)
\ee 
and obtain Eq.~\eqref{Eq:SHGsmallB}. We note that this asymptotic expression is valid provided not only $\omega_B \ll |\omega|, \, |\mu|/\hbar$, but also $\omega_B \ll \sqrt{|\omega^2 - 4\mu^2/\hbar^2|}$. To obtain accurate behavior in the vicinity of $\hbar |\omega| \approx 2 |\mu|$, one may use the Euler-Maclaurin formula.

Finally, in the low-frequency limit, $|\omega| \ll \min \left\{\omega_B, |\mu|/\hbar, \hbar \omega_B^2/|\mu|  \right\}$, we perform Taylor expansion and find that 
\be  
f_n(i\omega) \approx \frac{8(k_{n+1}-k_n)\omega^2}{\omega_B^4}.
\ee 
After plugging this expression into Eq.~\eqref{Eq:SHGgeneralxxz} we reproduce Eq.~\eqref{Eq:xxzloww}.

\subsection{${\boldsymbol \sigma^{xzx}}$ component of SHG}
\label{App:SHGxzx}

We continue our analysis with the component $\sigma^{xzx}(i\omega) = \sigma^{zxx}(i\omega) = [\chi^{xzx}(i\omega, i\omega) + \chi^{zxx}(i\omega, i\omega)]/ 2\omega^2$. The extra factor of 2 comes from the observation that the Kubo formula calculates the total contribution to $j^x(2\omega)$ from fields $E^x(\omega)$ and $E^z(\omega)$ in Eq.~\eqref{Eq:2ndOhm}, i.e., the left-hand side of Eq.~\eqref{Eq:sigmachi} should be understood in this case as $\sigma^{xzx} + \sigma^{zxx}$, and similarly for other SHG components. The whole calculation then is absolutely analogous to the previous section, even though lengthier, so we only highlight the key distinctive features. 

The sum of the form-factors for this component equals

\begin{align}
Z^{xzx}_{n_1 n_2 n_3} + Z^{zxx}_{n_1 n_2 n_3} = &-\frac{E_0}{4 E^2_{n_1} E_{n_3}} \delta_{n_1,n_2} \left[\delta_{|n_3|,|n_2|-1}(E_{n_3}+E_0)(E_{n_2}-E_0) + \delta_{|n_2|,|n_3|-1}(E_{n_2}+E_0)(E_{n_3}-E_0)  \right]  \nonumber \\ &-\frac{E_{n_1}^2-E_0^2}{4 E^2_{n_1} E_{n_3}} \delta_{n_1,-n_2} \left[\delta_{|n_2|,|n_3|-1}(E_{n_3}-E_0) - \delta_{|n_3|,|n_2|-1}(E_{n_3}+E_0)\right]  \nonumber \\ &- \frac{E_0}{4 E^2_{n_1} E_{n_2}} \delta_{n_1,n_3} \left[\delta_{|n_2|,|n_3|-1}(E_{n_2}+E_0)(E_{n_3}-E_0) + \delta_{|n_3|,|n_2|-1}(E_{n_3}+E_0)(E_{n_2}-E_0)  \right]  \nonumber \\ &-\frac{E_{n_1}^2-E_0^2}{4 E^2_{n_1} E_{n_2}} \delta_{n_1,-n_3} \left[\delta_{|n_3|,|n_2|-1}(E_{n_2}-E_0) - \delta_{|n_2|,|n_3|-1}(E_{n_2}+E_0)\right].
\end{align}
Plugging this into Eq.~\eqref{2ndOrderGeneric}, we obtain after some simplification

\begin{align}
\sigma^{xzx}(i\omega) &= -\frac{\eta e^4 v_F^3 B}{32 \pi^2 \hbar \omega^2}\int_{-\infty}^{\infty} d k_z
 \sum_{n_1, n_2} \frac{\Theta(\ve_2) - \Theta(\ve_1)}{(E_{n_2} - E_{n_1}+ i \hbar \omega)(E_{n_2} - E_{n_1}+ 2 i \hbar \omega)} \cdot \frac{E_0(E_{n_2}-E_{n_1})}{E_{n_1}^2 E_{n_2}^2} \times \nonumber \\ &\times \left[ \delta_{|n_1|, |n_2|-1} (E_{n_1} + E_0)(E_{n_2} - E_0) + \delta_{|n_2|, |n_1|-1} (E_{n_2} + E_0)(E_{n_1} - E_0) \right] + \nonumber \\ &+ \left[\Theta(\ve_2) - \Theta(\ve_1) \right] \left[ \frac1{\left(E_{n_2} + E_{n_1} - i\hbar\omega\right)\left(E_{n_2} - E_{n_1} + 2 i\hbar\omega\right)} +\frac1{\left(E_{n_2} + E_{n_1} + i\hbar\omega\right)\left(E_{n_2} - E_{n_1} - 2 i\hbar\omega\right)}  \right]  \times \nonumber\\ &\times \frac{ E_{n_2}^2 - E_0^2}{E_{n_1} E_{n_2}^2} \cdot \left[ \delta_{|n_2|, |n_1|-1} (E_{n_1} - E_0) - \delta_{|n_1|, |n_2|-1} (E_{n_1} + E_0)\right].
\end{align} 
This component is also an odd function of $\mu$ and $\eta$. Following the steps outlined in the previous section, we end up with Eq.~\eqref{Eq:SHGgeneralxzx} and all the limiting cases following from it.  

After analytic continuation, the result reads as
\begin{align}
    &\sigma^{xzx}(\omega) = \frac{\eta e^3}{32 \pi^2 \hbar^2}\frac{\omega_B^2}{\omega^2} \frac{\text{sgn}(\mu)}{\omega_B^4 - 4\omega^4} \left\{ \sum_{n=0}^{N_0(2\omega)}\left[(\omega_B^2 - 2\omega^2)(\omega_B^2 - 4\omega^2)-12 \omega^2 \omega_B^2 n  \right]f_n^-(2 \omega)  \right. \nonumber \\ &+\sum_{n=N_0(2\omega)+1}^{N_{\max}}\left[(\omega_B^2 - 2\omega^2)(\omega_B^2 - 4\omega^2)-12 \omega^2 \omega_B^2 n  \right]f_n^+(2 \omega) -\sum_{n=0}^{N_0(\omega)}\left[2(\omega_B^2 - \omega^2)(\omega_B^2 - 2\omega^2)-12 \omega^2 \omega_B^2 n  \right]f_n^{-}(\omega) \nonumber  \\ &\left.-  \sum_{n=N_0(\omega)+1}^{N_{\max}}\left[2(\omega_B^2 - \omega^2)(\omega_B^2 - 2\omega^2)-12 \omega^2 \omega_B^2 n  \right]f_n^{+}(\omega) \right\},
\end{align}
where $N_0(\omega)$ and $N_{\max}$ are defined in Eq.~\eqref{AppEq:defs} and $f_n^{\pm}(\omega)$ are defined in Eq.~\eqref{AppEq:fpm}.

\subsection{${\boldsymbol \sigma^{xzy}}$ component of SHG}
\label{App:SHGxzy}

Finally, we calculate the components $\sigma^{xzy}(i\omega) = \sigma^{zxy}(i\omega)$. While the basic approach for calculating these components is similar to the previous ones, certain technical details are rather nontrivial, so we discuss them in more detail. 

We find it more convenient to subtract the corresponding contribution at zero chemical potential, i.e., we define  

\be  
\sigma^{xzy}(i\omega) \equiv [\chi^{xzy}(i\omega, i\omega,\mu) + \chi^{zxy}(i\omega, i\omega,\mu) - \chi^{xzy}(i\omega, i\omega,0) - \chi^{zxy}(i\omega, i\omega,0)]/2\omega^2. 
\ee
This extra contribution, obviously, does not depend on $\mu$; consequently, it cancels out after summation over the nodes with opposite chiralities (since each component is proportional to the chirality of the node). 

The sum of the form-factors for this component equals

\begin{align}
Z^{xzy}_{n_1 n_2 n_3} + Z^{zxy}_{n_1 n_2 n_3} &= \frac{i E_0}{4 E^2_{n_1} E_{n_3}} \delta_{n_1,n_2} \left[\delta_{|n_3|,|n_2|-1}(E_{n_3}+E_0)(E_{n_2}-E_0) - \delta_{|n_2|,|n_3|-1}(E_{n_2}+E_0)(E_{n_3}-E_0)  \right]  \nonumber \\ &-i\frac{E_{n_1}^2-E_0^2}{4 E^2_{n_1} E_{n_3}} \delta_{n_1,-n_2} \left[\delta_{|n_2|,|n_3|-1}(E_{n_3}-E_0) + \delta_{|n_3|,|n_2|-1}(E_{n_3}+E_0)\right] + \nonumber \\ &+ \frac{i E_0}{4 E^2_{n_1} E_{n_2}} \delta_{n_1,n_3} \left[\delta_{|n_3|,|n_2|-1}(E_{n_3}+E_0)(E_{n_2}-E_0) - \delta_{|n_2|,|n_3|-1}(E_{n_2}+E_0)(E_{n_3}-E_0)  \right] + \nonumber \\ &+i\frac{E_{n_1}^2-E_0^2}{4 E^2_{n_1} E_{n_2}} \delta_{n_1,-n_3} \left[\delta_{|n_3|,|n_2|-1}(E_{n_2}-E_0) + \delta_{|n_2|,|n_3|-1}(E_{n_2}+E_0)\right].
\end{align}
The derivation then is lengthy but straightforward, and we arrive at Eq.~\eqref{Eq:SHGgeneralxzy}. This component is an even function of $\mu$ and odd function of $\eta$. 

The analytic continuation, however, requires extra care and better be performed directly in integrals from which Eq.~\eqref{Eq:SHGgeneralxzy} was derived, rather than from Eq.~\eqref{Eq:SHGgeneralxzy} itself. Below, we calculate the corresponding integrals in Matsubara frequencies and perform their analytic continuations explicitly.

\begin{flalign}
&1. \qquad F_{1,n}(i\omega) = \int_0^{k_n}\frac{dk}{\sqrt{k^2 + \omega_B^2 n}}\cdot \frac1{k^2 + \omega_B^2 n + \frac{\omega^2}4} = \frac2{\omega \sqrt{\omega^2 + 4 \omega_B^2 n}}\ln \frac{k_0\sqrt{\omega^2 + 4\omega_B^2 n}+\omega k_n}{k_0\sqrt{\omega^2 + 4\omega_B^2 n}-\omega k_n}.&&
\end{flalign}
This component is well defined for $n \ge 1$, and we used $k_n = \sqrt{(\mu/\hbar)^2 - \omega_B^2 n}$ for $n \le N_{\max}$ to simplify the answer (we note that $k_0 = |\mu|/\hbar$). Upon analytic continuation $i \omega \to \omega + i0$, this component becomes

\begin{align}
F_{1,n}(i\omega)  \quad \Longrightarrow \quad &F_{1,n}^+(\omega) \equiv \frac{4}{\omega \sqrt{4 \omega_B^2 n - \omega^2}} \arctan \left[\frac{k_n \omega}{k_0 \sqrt{4 \omega_B^2 n - \omega^2}}\right], \quad &&\text{for} \quad n > \frac{\omega^2}{4 \omega_B^2},  \nonumber\\ F_{1,n}(i\omega)  \quad \Longrightarrow \quad &F_{1,n}^-(\omega) \equiv\frac{2}{\omega \sqrt{\omega^2 - 4 \omega_B^2 n}} \left( \ln \left| \frac{k_0\sqrt{\omega^2 - 4 \omega_B^2 n} - \omega k_n}{k_0{\sqrt{\omega^2 - 4 \omega_B^2 n} + \omega k_n}}  \right| +\pi i \Theta\left(2k_0 - |\omega|\right)  \right),  \quad &&\text{for} \quad n < \frac{\omega^2}{4 \omega_B^2}. 
\end{align}

\begin{flalign}
&2. \qquad F_{2,n}(i\omega) = \int_0^{k_n}\frac{dk}{\sqrt{k^2 + \omega_B^2 n}}\cdot \frac1{k^2 + a_n^2(i\omega)} = \frac{\omega}{a_n(i\omega)(\omega^2 + \omega_B^2)}\ln \frac{2 \omega k_0 a_n(i\omega) + k_n (\omega^2 + \omega_B^2)}{2 \omega k_0 a_n(i\omega) - k_n (\omega^2 + \omega_B^2)},&&
\end{flalign}
which is, again, well-defined for $n \ge 1$. The analytic continuation reads

\begin{align} 
F_{2,n}(i\omega)  \quad \Longrightarrow \quad &F_{2,n}^+(\omega) \equiv \frac{2\omega}{a_n(\omega) (\omega_B^2 - \omega^2)}\arctan \left[ \frac{k_n(\omega_B^2 - \omega^2)}{2 k_0 \omega a_n(\omega)}  \right], \quad &&\text{for} \quad n > \frac{(\omega_B^2 - \omega^2)^2}{4 \omega_B^2 \omega^2}, \nonumber\\ F_{2,n}(i\omega)  \quad \Longrightarrow \quad &F_{2,n}^-(\omega) \equiv\frac{\omega}{|a_n(\omega)|(\omega_B^2 - \omega^2)}\ln \left| \frac{2 \omega k_0 |a_n(\omega)| - k_n(\omega_B^2 - \omega^2)}{2 \omega k_0 |a_n(\omega)| + k_n(\omega_B^2 - \omega^2)} \right| -  \\ &\frac{\pi i \omega}{|a_n(\omega)|(\omega_B^2 - \omega^2)} \Theta\left(2 k_0 |\omega| - |\omega_B^2 - \omega^2|\right),  \quad &&\text{for} \quad n < \frac{(\omega_B^2 - \omega^2)^2}{4 \omega_B^2 \omega^2}. \nonumber
\end{align}
Here 
\be  
a_n(\omega) = \sqrt{\omega_B^2 n - \frac{(\omega_B^2 - \omega^2)^2}{4 \omega^2}},
\ee
which is purely real for $n > (\omega_B^2 - \omega^2)^2/4\omega^2 \omega_B^2$, and 
\be  
|a_n(\omega)| = \sqrt{\frac{(\omega_B^2 - \omega^2)^2}{4 \omega^2} - \omega_B^2 n}
\ee
for $n < (\omega_B^2 - \omega^2)^2/4\omega^2 \omega_B^2$.

\begin{flalign}
&3. \qquad F_{3,n}(i\omega) = \int_0^{k_{n+1}}\frac{dk}{\sqrt{k^2 + \omega_B^2 (n+1)}}\cdot \frac1{k^2 + a_{n}^2(i\omega)} = \frac{\omega}{a_n(i\omega)(\omega^2 - \omega_B^2)}\ln \frac{2 \omega k_0 a_n(i\omega) + k_{n+1} (\omega^2 - \omega_B^2)}{2 \omega k_0 a_n(i\omega) - k_{n+1} (\omega^2 - \omega_B^2)},&&
\end{flalign}
for all $n \ge 0$. The analytic continuation reads

\begin{align} 
F_{3,n}(i\omega)  \quad \Longrightarrow \quad &F_{3,n}^+(\omega) \equiv \frac{2\omega}{a_n(\omega) (\omega_B^2 + \omega^2)}\arctan \left[ \frac{k_{n+1}(\omega_B^2 + \omega^2)}{2 k_0 \omega a_n(\omega)}  \right], \quad &&\text{for} \quad n > \frac{(\omega_B^2 - \omega^2)^2}{4 \omega_B^2 \omega^2}, \nonumber\\ F_{3,n}(i\omega)  \quad \Longrightarrow \quad &F_{3,n}^-(\omega) \equiv\frac{\omega}{|a_n(\omega)|(\omega_B^2 + \omega^2)}\ln \left| \frac{2 \omega k_0 |a_n(\omega)| - k_{n+1}(\omega_B^2 + \omega^2)}{2 \omega k_0 |a_n(\omega)| + k_{n+1}(\omega_B^2 + \omega^2)} \right| -  \\ &\frac{\pi i \omega \, \text{sgn}(\omega_B^2 - \omega^2)}{|a_n(\omega)|(\omega_B^2 + \omega^2)} \Theta\left(2 k_0 |\omega| - |\omega_B^2 + \omega^2|\right),  \quad &&\text{for} \quad n < \frac{(\omega_B^2 - \omega^2)^2}{4 \omega_B^2 \omega^2}. \nonumber
\end{align}

Finally, we have:

\begin{flalign}
&4. \quad \int_0^{\mu^2}\frac{dx}{x + \hbar^2 a_0^2(i\omega)} = \ln \frac{\mu^2 + \hbar^2 a_0^2(i\omega)}{\hbar^2 a_0^2(i\omega)} \quad \Longrightarrow \quad \ln \left| \frac{\hbar^2|a_0(\omega)|^2-\mu^2}{\hbar^2 |a_0(\omega)|^2}  \right|  - \pi i \text{sgn} \left[ \omega(\omega_B^2 - \omega^2) \right] \Theta\left[\frac{\mu^2}{\hbar^2} - |a_0(\omega)|^2  \right], &&
\end{flalign}
with $|a_0(\omega)| = \left|\omega_B^2 - \omega^2\right|/2|\omega|$.

Noticing that 

\be  
g_n(i\omega) = \frac{(\omega_B^2 - \omega^2)}{\omega}F_{3,n}(i\omega) - \frac{(\omega_B^2 + \omega^2)}{\omega}F_{2,n}(i\omega),
\ee 
where $g_n(i\omega)$ is defined in Eq.~\eqref{Eq:gn}, and collecting everything together, we obtain

\begin{align} \label{AppEq:sigmaxzyreal}
&\sigma^{xzy}(\omega) = i\frac{\eta e^3}{8\pi^2 \hbar^2} \cdot \frac{\omega_B^2}{\omega(2\omega^2 + \omega_B^2)} \left\{\ln\left|\frac{(\omega_B^2 - \omega^2)^2-4\omega^2 k_0^2}{(\omega_B^2 - 4\omega^2)^2-16\omega^2 k_0^2} \cdot \frac{(\omega_B^2 - 4\omega^2)^2}{(\omega_B^2 - \omega^2)^2} \cdot \frac{k_0(\omega_B^2 - \omega^2)+ k_1(\omega_B^2 + \omega^2)}{k_0(\omega_B^2 - \omega^2)- k_1(\omega_B^2 + \omega^2)}\right|  \right. \nonumber \\
&+\ln\left|\frac{k_0(\omega_B^2 - 4\omega^2)- k_1(\omega_B^2 + 4\omega^2)}{k_0(\omega_B^2 - 4\omega^2)+ k_1(\omega_B^2 + 4\omega^2)}\right|    + \pi i \, \text{sgn}(\omega)\left[ \text{sgn}\left(\omega_B^2 - 4 \omega^2\right) \cdot \Theta\left(4|\omega|k_0 - |\omega_B^2 - 4\omega^2|\right) \right. \nonumber \\ &\left. \left.  - \Theta\left(4|\omega|k_0 - |\omega_B^2 + 4\omega^2|\right)  - \text{sgn}\left(\omega_B^2 - \omega^2\right) \cdot \Theta\left(2|\omega|k_0 - |\omega_B^2 - \omega^2|\right) + \Theta\left(2|\omega|k_0 - |\omega_B^2 + \omega^2|\right) \right] \right\}  \nonumber \\  &-i \frac{3\eta e^3}{16 \pi^2 \hbar^2} \cdot \frac{\omega_B^2}{\omega(4\omega^4 - \omega_B^4)} \sum_{n=1}^{N_{\max}} \left\{ 4n \omega_B^2 \omega^2 F_{1,n}(\omega) \right. \nonumber \\ &- \frac{(\omega_B^2 - \omega^2)(\omega_B^2 - 2\omega^2) - 6\omega^2 \omega_B^2 n}{3\omega^2} \left[(\omega_B^2 + \omega^2)F_{3,n}(\omega) - (\omega_B^2 - \omega^2)F_{2,n}(\omega)  \right]  \nonumber \\ &\left. + \frac{(\omega_B^2 - 2\omega^2)(\omega_B^2 - 4\omega^2) - 12\omega^2 \omega_B^2 n}{12\omega^2} \left[(\omega_B^2 + 4\omega^2)F_{3,n}(2\omega) - (\omega_B^2 - 4\omega^2)F_{2,n}(2\omega)  \right]\right\}.
\end{align}
In this expression, summation over $n$ should be understood in the following sense:

\begin{align} 
&\sum_{n=1}^{N_{\max}}n F_{1,n}(\omega) = \sum_{n=1}^{\tilde N_{0}(\omega)}n F_{1,n}^-(\omega) + \sum_{n=\tilde N_{0}(\omega) + 1}^{N_{\max}}n F_{1,n}^+(\omega), \qquad  \tilde N_{0}(\omega) = \min \left\{ \left\lfloor \frac{\omega^2}{4\omega_B^2} \right\rfloor , N_{\max} \right\}, \nonumber \\ &\sum_{n=1}^{N_{\max}}F_{(2,3),n}(\omega) = \sum_{n=1}^{N_{0}(\omega)}F_{(2,3),n}^-(\omega) + \sum_{n=N_{0}(\omega) + 1}^{N_{\max}}F_{(2,3),n}^+(\omega), \nonumber \\ &\sum_{n=1}^{N_{\max}}n F_{(2,3),n}(\omega) = \sum_{n=1}^{N_{0}(\omega)}n F_{(2,3),n}^-(\omega) + \sum_{n=N_{0}(\omega) + 1}^{N_{\max}}n F_{(2,3),n}^+(\omega), \nonumber \\ &\sum_{n=1}^{N_{\max}}F_{(2,3),n}(2\omega) = \sum_{n=1}^{N_{0}(2\omega)}F_{(2,3),n}^-(2\omega) + \sum_{n=N_{0}(2\omega) + 1}^{N_{\max}}F_{(2,3),n}^+(2\omega), \nonumber \\ &\sum_{n=1}^{N_{\max}}n F_{(2,3),n}(2\omega) = \sum_{n=1}^{N_{0}(2\omega)}n F_{(2,3),n}^-(2\omega) + \sum_{n=N_{0}(2\omega) + 1}^{N_{\max}}n F_{(2,3),n}^+(2\omega), 
\end{align}
where $N_0(\omega)$ and $N_{\max}$ are defined in Eq.~\eqref{AppEq:defs}.

In the EQL, $\omega_B > k_0$, $N_{\max} = 0$, so only $n=0$ term contributes. Furthermore, since $k_1 = 0$ in this limit, we find that Eq.~\eqref{AppEq:sigmaxzyreal} reduces to Eq.~\eqref{Eq:sigmaxzyEQL}.

To derive the asymptotic behavior at small magnetic fields $\omega_B \ll |\omega|$, $k_0$, $\sqrt{|\omega^2 - 4k_0^2|}$, $\sqrt{|\omega^2 - k_0^2|}$, we employ the Euler-Maclaurin formula and switch from summation to integration: 

\be  \label{AppEq:EM}
\sum_{i=m}^n f(i) = \int_m^n f(x) dx + \frac{f(m) + f(n)}2  + \frac16 \frac{f'(n) - f'(m)}{2!} + \ldots
\ee 
Treating the terms $n=0$ and $n=N_{\max}$ separately, we apply Eq.~\eqref{AppEq:EM} to the rest of the terms in Eq.~\eqref{Eq:SHGgeneralxzy}, $n=1,\ldots,N_{\max}-1$. Carefully collecting all the contributions of the order $\omega_B^3$ and $\omega_B^4$, we find

\begin{align} 
\sigma^{xzy}(i\omega) \approx &\frac{3 \eta e^3}{8 \pi^2 \hbar^2} \cdot \frac{\hbar^3 \omega_B^3 |\mu|}{\omega(\hbar^2 \omega^2 + \mu^2)(\hbar^2 \omega^2 + 4\mu^2)} \zeta\left( -\frac12, \frac{\mu^2}{\hbar^2 \omega_B^2}- \left\lfloor  \frac{\mu^2}{\hbar^2 \omega_B^2} \right\rfloor \right) - \\  &\frac{\eta e^3}{64 \pi^2 \hbar^2}\cdot \frac{\omega_B^4}{\omega^5} \cdot \left[ \frac{5 \hbar^4 \omega^4 + 16 \hbar^2 \omega^2 \mu^2 + 48\mu^4}{(\hbar^2 \omega^2 + 4\mu^2)^2} + \frac23\ln \frac{2^6 \omega^6 \mu^6 (\hbar^2 \omega^2 + \mu^2)}{\omega_B^6(\hbar^2 \omega^2 + 4\mu^2)^4}\right], \nonumber
\end{align}
where $\zeta(-1/2,x)$ is the Hurwitz zeta function defined through the identity
\be 
\sum_{n=0}^N \sqrt{n+x} = \zeta\left(-\frac12,x\right) - \zeta\left(-\frac12,x+N+1\right). 
\ee
At vanishing field, the first term $\propto \omega_B^3$ is formally the dominant one, and the Hurwitz zeta function leads to pronounced oscillations. However, the second term, which is of the order of $\mathcal{O}(\omega_B^4)$, quickly becomes important as the magnetic field increases, so we keep it as well. 

We note that the series in Eq.~\eqref{AppEq:EM} is the asymptotic one, so we cut it neglecting the boundary terms with higher derivatives, which are formally $\propto \omega_B^4/\omega^5$. We checked, however, that the contribution of these terms is numerically negligible; furthermore, since these terms do not depend on the chemical potential $\mu$, they cancel out after summation over the nodes with opposite chiralities. 

After performing analytic continuation, we obtain Eq.~\eqref{Eq:sigmaxzysmallB}. The analytic continuation of the logarithm is more conveniently performed from the corresponding integral representations, analogously to how it has been done when deriving Eq.~\eqref{AppEq:sigmaxzyreal}, i.e.,

\begin{align}
   &\ln \frac{\hbar^6 \omega^6(\hbar^2 \omega^2 + \mu^2)}{(\hbar^2 \omega^2 + 4\mu^2)^4} = \int_0^{\mu^2}\frac{dx}{x+ \hbar^2 \omega^2} - 4\int_0^{4\mu^2}\frac{dx}{x+\hbar^2 \omega^2} \quad \Longrightarrow \quad \int_0^{\mu^2}\frac{dx}{x -  \hbar^2 (\omega+i0)^2} - 4\int_0^{4\mu^2}\frac{dx}{x-\hbar^2 (\omega+i0)^2} =  \nonumber \\  &\ln \left|\frac{\hbar^6 \omega^6(\hbar^2 \omega^2 - \mu^2)}{(\hbar^2 \omega^2 - 4\mu^2)^4}\right| + \pi i \, \text{sgn}(\omega) \left[\Theta\left(\mu^2 - \hbar^2 \omega^2  \right) -  4\Theta\left(4\mu^2 - \hbar^2 \omega^2  \right) \right].
\end{align}
We also used the identities
\begin{align}
    &\frac1{(\mu^2 + \hbar^2 \omega^2)^2} = -\frac{\partial}{\partial \mu^2} \left(\frac1{\mu^2 + \hbar^2 \omega^2}\right) \Longrightarrow  -\frac{\partial}{\partial \mu^2} \left[\frac1{\mu^2 - \hbar^2 \omega^2} + \pi i\, \text{sgn}(\omega) \delta\left(\mu^2 - \hbar^2 \omega^2   \right)  \right] = \nonumber \\ & \frac1{(\mu^2 - \hbar^2 \omega^2)^2} - \pi i \, \text{sgn}(\omega) \delta'\left( \mu^2 - \hbar^2 \omega^2 \right), \\ &\left(5 \hbar^4 \omega^4 - 16 \hbar^2 \omega^2 \mu^2 + 48 \mu^4\right) \delta'\left(4\mu^2 - \hbar^2 \omega^2  \right) = -4\mu^2 \left[\delta'\left(\hbar|\omega| - 2|\mu|  \right) - \frac{\delta\left(\hbar|\omega| - 2|\mu|  \right)}{|\mu|}  \right]. \nonumber
\end{align}

To derive low-frequency behavior, $\omega \to 0$, we expand Eq.~\eqref{Eq:SHGgeneralxzy} to the first nonvanishing (linear) power of $\omega$, focusing on the case $\hbar \omega_B < |\mu|$ first. When doing so, we consider terms $n=0$ and $n=N_{\max}$ separately, and obtain

\begin{align}\label{AppEq:sigmaxzyloww}
\sigma^{xzy}(i\omega) &\approx - \frac{3\eta e^3 \omega|\mu|}{2\pi^2 \hbar^2 \omega_B^2 (|\mu| + \hbar k_1)} - \frac{3\eta e^3 \omega}{4\pi^2 \hbar^3 \omega_B^4 |\mu|}\sum_{n=1}^{N_{\max}-1}\left[ \hbar^2 \omega_B^2 k_n + 2\mu^2(k_n - k_{n+1}) \right] - \frac{3\eta e^3 \omega k_{N_{\max}} (2\mu^2 + \omega_B^2 \hbar^2)}{4\pi^2 \hbar^3 \omega_B^4 |\mu|} = \nonumber \\ &= -\frac{3 \eta e^3 \omega}{4\pi^2 \hbar^2 \omega_B^2} \left(\frac{2\mu^2}{\hbar^2 \omega_B^2} + \frac{\hbar}{|\mu|} \sum_{n=1}^{N_{\max}} k_n  \right),
\end{align}
which after analytic continuation reproduces Eq.~\eqref{Eq:sigmaxzyloww}. Finally, applying low-frequency expansion to the EQL ($\hbar \omega_B > |\mu|$) expression given by  Eq.~\eqref{Eq:sigmaxzyEQL}, we find that Eqs.~\eqref{Eq:sigmaxzyloww} and~\eqref{AppEq:sigmaxzyloww} are valid in this case as well. 

\section{Semiclassical calculation \label{App:semiclassics}}

We start the analysis in this Appendix with the semiclassical equations of motion which have the form: 

\begin{align}
\hbar \dot \br  &= \nabla_\bk \ve_\bk - \hbar \dot \bk \times {\bf \Omega}_{\bk}, \nonumber \\ 
\hbar \dot \bk &= -e \bE - e \dot\br \times \bB,
\end{align}
where ${\bf \Omega}_\bk = i \langle \nabla_{\bk} u_{\bk} | \times \nabla_{\bk} u_{\bk}    \rangle$ is the Berry curvature. In the presence of external magnetic field $\bB$, the quasiparticle energy dispersion is modified according to 
\be
\ve_\bk = \ve^0_\bk - {\bf m}_\bk \cdot \bB,
\ee 
where $\ve^0_\bk$ is the bare band energy at $\bB=0$, $H_\bk |u_\bk\rangle = \ve^0_\bk |u_\bk\rangle$, and the orbital magnetic moment is given by

\be 
{\bf m}_\bk = -i \frac{e}{2\hbar} \langle \nabla_{\bk} u_{\bk} | \times (H_\bk - \ve_\bk^0)| \nabla_{\bk} u_{\bk}    \rangle .
\ee

These equations can be readily resolved to give

\begin{align}
\dot \br &= \frac1{\hbar D_\bk} \left\{ \nabla_\bk \ve_\bk + e \bE \times {\bf \Omega_\bk}  + \frac{e}{\hbar} \bB \left( {\bf \Omega_\bk} \cdot \nabla_\bk \ve_\bk  \right) \right\}, \nonumber \\ \dot \bk &= \frac1{\hbar D_\bk}\left\{  -e \bE - \frac{e}{\hbar} \nabla_\bk \ve_\bk \times \bB - \frac{e^2}{\hbar} (\bE \cdot \bB) {\bf \Omega_\bk}  \right\}, \label{AppEq:EOMresolved}
\end{align} 
where we have also introduced the phase-space correction factor $D_\bk = 1 + (e/\hbar) (\bB \cdot {\bf \Omega_\bk}).$

The contribution of the magnetization current $-e \int_{\bk}\nabla_\br \times {\bf m}_\bk \cdot f$ can be neglected in the case of a uniform system. Then, the overall current density is given by 

\be  
{\bf j} = -e \int \frac{d^3k}{(2\pi)^3} D_\bk \dot \br f = - \frac{e}{\hbar} \int \frac{d^3k}{(2\pi)^3} \left[ \nabla_\bk \ve_\bk + e \bE\times{\bf \Omega}_\bk + \frac{e}{\hbar} \bB (\nabla_\bk \ve_\bk \cdot {\bf \Omega}_\bk) \right] f. \label{SMEq:jtotsc}
\ee 
The distribution function $f$ satisfies the conventional kinetic equation, which for the uniform system, in the relaxation time approximation, takes the form:

\be 
\frac{\partial f}{\partial t} + \dot \bk \nabla_\bk f = - \frac{f-f_0}{\tau},
\ee 
where $\dot \bk$ is given by Eq.~\eqref{AppEq:EOMresolved}, $\tau$ is the relaxation time, and $f_0=f_0(\ve_\bk)$ is the equilibrium Fermi-Dirac distribution (as a function of $\ve_\bk$, not $\ve_\bk^0$), which at zero temperature takes the form $f_0 = \Theta(\mu - \ve_\bk)$. In the case of a periodic electric field, 
\be
\bE(t) = \bE_\omega e^{-i\omega t} + \bE_{\omega}^* e^{i\omega t},
\ee 
the steady-state distribution function admits the expansion 

\be  
f(t) = f_0 + \delta f_0 + (f_1 e^{-i \omega t} +  f_1^* e^{i \omega t}) + (f_2 e^{-2i \omega t} + f_2^* e^{2i \omega t}) + \ldots,
\ee 
where $f_1 \propto E = |\bE_\omega|$, $f_2 \propto E^2$, etc. The time-independent correction $\delta f_0 \propto E^2$ appears due to the second-order optical transitions and contributes to PGE (second-order dc current). The reality condition requires that $\bE_{-\omega} = \bE_{\omega}^*$. Expanding the kinetic equation into different harmonics, we find a set of coupled equations:
\begin{align}
D_\bk \left(i\omega - \frac1\tau\right)  f_1 &= -  \frac{e}{\hbar}\left[ \bE_\omega + \frac{e}{\hbar} (\bE_\omega \cdot \bB){\bf \Omega} \right]\cdot \nabla_\bk f_0 - \frac{e}{\hbar^2} \nabla_\bk f_1 \cdot \nabla_\bk \ve_\bk \times \bB, \nonumber \\ 
D_\bk \left(2i\omega - \frac1\tau\right)  f_2 &= -  \frac{e}{\hbar}\left[ \bE_\omega + \frac{e}{\hbar} (\bE_\omega \cdot \bB){\bf \Omega} \right]\cdot \nabla_\bk f_1 - \frac{e}{\hbar^2} \nabla_\bk f_2 \cdot \nabla_\bk \ve_\bk \times \bB, \\ - D_\bk \frac1\tau  \delta f_0 &= -  \frac{e}{\hbar}\left[ \bE_\omega^* + \frac{e}{\hbar} (\bE_\omega^* \cdot \bB){\bf \Omega} \right]\cdot \nabla_\bk f_1 - \frac{e}{2\hbar^2} \nabla_\bk \delta f_0 \cdot \nabla_\bk \ve_\bk \times \bB + \text{c.c.} \nonumber
\end{align}

Generically, this is a complicated system of differential equations. However, it admits a straightforward iterative solution in the case of a small magnetic field. Up to the linear order in $\bB$, the solution reads

\begin{align} \label{AppEq:f_isol}
f_1 &= \frac{e \tau}{\hbar D_\bk (1 - i\omega\tau)} \left\{   \left[ \bE_\omega + \frac{e}{\hbar} (\bE_\omega \cdot \bB){\bf \Omega}_\bk \right]\cdot \nabla_\bk f_0 + \frac{e \tau}{\hbar^2(1- i \omega\tau)}  [\nabla_\bk \ve_\bk \times \bB] \cdot \nabla_\bk \left( \bE_\omega \cdot \nabla_\bk f_0 \right)   \right\}, \nonumber \\ 
f_2 &= \frac{e \tau}{\hbar D_\bk (1 - 2i\omega\tau)} \left\{   \left[ \bE_\omega + \frac{e}{\hbar} (\bE_\omega \cdot \bB){\bf \Omega}_\bk \right]\cdot \nabla_\bk f_1 + \frac{e \tau}{\hbar^2(1- 2i \omega\tau)}  [\nabla_\bk \ve_\bk \times \bB] \cdot \nabla_\bk \left( \bE_\omega \cdot \nabla_\bk f_1 \right)   \right\}, \\  \delta f_0 &= \frac{e \tau}{\hbar D_\bk} \left\{   \left[ \bE_\omega^* + \frac{e}{\hbar} (\bE_\omega^* \cdot \bB){\bf \Omega}_\bk \right]\cdot \nabla_\bk f_1 + \frac{e \tau}{\hbar^2}  [\nabla_\bk \ve_\bk \times \bB] \cdot \nabla_\bk \left( \bE_\omega^* \cdot \nabla_\bk f_1 \right)   \right\}  + \text{c.c.} \nonumber
\end{align}
We stress here that the equations determining $f_1$ and $f_2$ are well defined in the limit $\tau \to \infty$ and lead to the solutions which are still described by Eq.~\eqref{AppEq:f_isol}, i.e., the limits $\bB \to 0$ and $\tau \to \infty$ commute for these harmonics. On the contrary, the equation for $\delta f_0$ becomes much more complicated if we set $\tau \to \infty$ from the beginning, and it does not allow for a solution then which is simply an expansion in powers of $\bB$. This implies that Eq.~\eqref{AppEq:f_isol} is not applicable for $\delta f_0$ in the limit $\tau \to \infty$  any longer, indicating that the most general solution depends sensitively on the parameter $\omega_B \tau$.  

Having solved for the distribution function in the limit  $\bB \to 0$ (at fixed finite $\tau$), we plug the solution given by Eq.~\eqref{AppEq:f_isol} into Eq.~\eqref{SMEq:jtotsc} to calculate the nonlinear current. Its quadratic part  is given by the sum of the dc current $\bj_2^{\text{dc}}$ (photocurrent) and the SHG current $\bj_2^{2\omega} e^{-2i\omega t} + \text{c.c.}$:

\begin{align} \label{AppEq:2ndcurrents}
\bj_2^{\text{dc}} &= -\frac{e}{\hbar} \int \frac{d^3k}{(2\pi)^3}\left\{\left[  \nabla_\bk \ve_\bk + \frac{e}{\hbar}(\nabla_\bk \ve_\bk \cdot \bOmega_\bk)\bB   \right]\delta f_0 + e \bE_\omega\times \bOmega_\bk f_1^* + e \bE_\omega^*\times \bOmega_\bk f_1\right\}, \nonumber \\ \bj_2^{2\omega}  &= -\frac{e}{\hbar} \int \frac{d^3k}{(2\pi)^3}\left\{\left[  \nabla_\bk \ve_\bk + \frac{e}{\hbar}(\nabla_\bk \ve_\bk \cdot \bOmega_\bk)\bB   \right]f_2 + e \bE_\omega\times \bOmega_\bk f_1  \right\}.
\end{align}
Finally, we apply these general equations to the case of a Weyl node, which is characterized by
\be  
\ve^0_\bk = \zeta v_F \hbar k, \qquad {\bf \Omega}_\bk = - \frac{\zeta \eta}{2 k^3} \bk, \qquad {\bf m}_\bk = - \eta \frac{e v_F}{2k^2} \bk, 
\ee 
where $\eta = \pm 1$ is the chirality of the node and $\zeta = +1/-1$ corresponds to the conduction/valence band. We assume that the electric and magnetic fields are given by $\bE_\omega = (E_x,E_y,E_z)^T$ and $\bB = (0,0,B)^T$, correspondingly. Expanding Eqs.~\eqref{AppEq:2ndcurrents} up to linear order in $B$ (recall that $f_0$ depends on $\ve_\bk = \ve_\bk^0 - {\bf m}_\bk \cdot \bB$ rather than on $\ve_\bk^0$), we evaluate integrals over $\bk$ and find
\begin{align}
\bj_2^{2\omega} \approx   &\frac{\eta e^4 B v_F^2}{12 \pi^2 \hbar^2 \mu} \cdot \frac{\tau^2(2-3i\omega\tau)}{(1-i\omega\tau)^2(1-2i\omega\tau)} \left( \begin{matrix}E_x E_z \\ E_y E_z \\ -(E_x^2 + E_y^2) \end{matrix}   \right), \nonumber \\
\bj_2^{\text{dc}} \approx  &i\frac{ \eta e^3}{6 \pi^2 \hbar^2} \cdot \frac{\tau^2 \omega}{1+\omega^2\tau^2} \bE_\omega \times \bE^*_\omega  +  \\ &\frac{\eta e^4 B v_F^2}{12 \pi^2 \hbar^2 \mu} \cdot \frac{\tau^2}{(1+\omega^2\tau^2)^2} \left\{ \left( \begin{matrix}
    2(E_x E_z^* + E_x^* E_z) \\ 2(E_y E_z^* + E_y^* E_z) \\ -4(|E_x|^2 + |E_y|^2)
\end{matrix}   \right)  - i \omega \tau (1-\omega^2 \tau^2) \left( \begin{matrix}
    E_z E_x^* - E_z^* E_x \\ E_z E_y^* - E_z^* E_y \\ 0
\end{matrix}   \right)  \right\}. \nonumber
\end{align}
The part linear in $B$ exactly corresponds to Eqs.~\eqref{Eq:semidc}--\eqref{Eq:semishg}. The contribution to photocurrent independent of $B$, surprisingly, does not follow from our exact microscopic result at $\bB=0$, Eq.~\eqref{SMEq:chi_B=0}. This observation poses a separate question about the origin of the discrepancy in the answers obtained within different approaches. One possible resolution could be the proper gauge choice for the electric field. Indeed, since this term does not depend on chemical potential, it necessarily cancels out after summation over the nodes with the opposite chiralities and, consequently, does not contribute to the physical answer. Another opportunity might be the inclusion of both conduction and valence bands into  semiclassical calculation and accounting for the interband transitions. We leave the detailed analysis of this issue to future work.

\twocolumngrid
\bibliography{SecondOrderResponseRefs_4}

\end{document}